\documentclass[journal=achre4,manuscript=article,articletitle=true,layout=twocolumn,etalmode=truncate,maxauthors=50]{achemso}

\usepackage[version=3]{mhchem} 
\usepackage{lipsum}
\usepackage{amssymb}
\usepackage{amsmath}
\usepackage{amsfonts}
\usepackage{dcolumn}
    \newcolumntype{d}[1]{D{.}{.}{#1}}
\usepackage{graphics,latexsym} 
\usepackage{subcaption}
\usepackage{multicol}
\usepackage{float}
\usepackage{longtable}

\usepackage[margin=10pt,font=small,labelfont=bf,   justification=justified,format=plain]{caption} 
\usepackage{eucal}
\usepackage[english]{babel}
\usepackage[usenames, dvipsnames]{color}
\usepackage[dvipsnames]{xcolor}
\usepackage[perpage]{footmisc}

\newcommand{\me}{\mathrm{e}}

\usepackage[intoc]{nomencl} 
\usepackage{ifpdf} 

\usepackage{array}   
\newcolumntype{P}[1]{>{\centering\arraybackslash}p{#1}}

\usepackage{appendix}

\usepackage{multirow}
\usepackage{breqn}
\usepackage{soul}
\usepackage{ulem}

\usepackage{mathtools}

\usepackage{dcolumn}
\usepackage{xcolor}
\usepackage{diagbox}

\setkeys{acs}{etalmode=truncate,maxauthors=0}

\author{Diego T. Gomez}
\affiliation{Department of Chemical \& Biomolecular Engineering, Tulane University, New Orleans LA 70118}
\email{dgomez1@tulane.edu}

\author{Lawrence R. Pratt}
\affiliation{Department of Chemical \& Biomolecular Engineering, Tulane University, New Orleans LA 70118}
\email{lpratt@tulane.edu}

\author{Dilipkumar N. Asthagiri}
\affiliation{Department of Chemical and Biomolecular Engineering, Rice University, Houston TX 77005}
\email{dna6@rice.edu}

\author{Susan B. Rempe}
\affiliation{Center for Integrated Nanotechnologies, Sandia National Laboratories, Albuquerque NM 87185}
\email{slrempe@sandia.gov}

\title{Hydrated anions: From clusters to bulk solution with quasi-chemical theory}

\begin{document}

\section*{Conspectus}
The interactions of hydrated ions with molecular and
macromolecular solution and interface partners are strong on a chemical energy scale. 
 Here, we recount the foremost \textit{ab initio theory}  for evaluation
of hydration free energies of ions; namely,
\textit{quasi-chemical theory} (QCT).   We focus  on anions,
particularly halides but also the hydroxide anion, since
they have been outstanding challenges for all theories. 
 For example, this work supports understanding
the high selectivity for F$^{-}$ over Cl$^-$ in
fluoride-selective ion channels,
despite the identical charge and the size similarity of these ions.  QCT is built by
identification of inner-shell clusters, separate treatment of those
clusters, then integration of those results into the broader-scale
solution environment. Recent work has focused on a close comparison with
mass-spectrometric measurements of ion-hydration equilibria.  We
delineate how \textit{ab initio} molecular dynamics (AIMD) calculations
on ion-hydration clusters, elementary statistical thermodynamics, and
electronic structure calculations on cluster structures sampled from the
AIMD calculations obtain just the free energies extracted from the
cluster experiments.    That theory-experiment comparison has not been
attempted before the work discussed here, but the agreement is excellent with
moderate computational effort.   This agreement reinforces both theory and
experiment, and provides a numerically accurate inner-shell contribution
to QCT.  The inner-shell complexes involving heavier halides
display strikingly 
asymmetric hydration clusters.  
Asymmetric hydration structures can be
problematic for evaluation 
of the QCT outer-shell contribution with the 
polarizable continuum model (PCM).   Nevertheless, QCT 
provides a favorable 
setting for exploitation of PCM when 
the inner-shell material shields the 
ion from the outer  solution 
environment. For the more asymmetrically hydrated, and thus less effectively 
shielded, heavier halide ions clustered with waters, the PCM is 
less satisfactory.  We therefore 
investigate an inverse procedure in which 
the inner-shell structures are sampled from 
readily available AIMD calculations on the 
bulk solutions.  This inverse procedure
is a remarkable improvement; our 
final results are in close agreement with 
a standard tabulation of hydration free energies, and the 
final composite 
results are independent of 
coordination number on the chemical energy scale of relevance, as they should be. Finally, comparison of anion hydration structure in clusters and bulk solutions from AIMD simulations emphasize some differences: the asymmetries of bulk solution 
inner-shell structures are moderated compared with clusters, but still present; and inner hydration shells fill to slightly higher average coordination numbers in bulk solution than in clusters.

\section*{Key References}
\begin{itemize}
\item Asthagiri,  D.;   Dixit,  P.;   Merchant,  S.;Paulaitis,   M.;    Pratt,   L.;    Rempe,   S.; Varma,  S.  Ion  Selectivity  from Local  Configurations of Ligands in Solutions and Ion  Channels. \textit{Chem. Phys. Lett.} \textbf{2010}, 485, 1–7.  This article\cite{Asthagiri:2010tj} gives a 
basic discussion of quasi-chemical theory (QCT), including some history, physical motivation, 
and the connection between \textit{direct}
and \textit{cluster} QCT.

\item Gomez, D. T.; Pratt, L. R.; Rogers, D. M.; Rempe,  S.  B.  Free  Energies  of  Hydrated Halide  Anions:   High  Through-Put  Computations   on   Clusters   to   Treat   Rough Energy-Landscapes.
\textit{Molecules} \textbf{2021}, 26, 3087.
This paper\cite{gomez2021free} details 
the theory and calculation of the QCT inner-shell
contributions, which lay the basis for testing
against the cluster-experimental association 
free energies.

\item Muralidharan,  A.;  Pratt,  L.  R.;  Chaudhari,  M.  I.;  Rempe,  S.  B.  Quasi-chemical Theory for Anion Hydration and Specific Ion Effects:  Cl$^-$(aq)  vs.  F$^-$(aq). \textit{Chem. Phys. Letts.: X} \textbf{2019}, 4, 100037. This paper\cite{ajay_cl} shows how to approximate
the outer-shell contribution by combining the PCM
model with structure sampling from dynamical
cluster simulations.

\item  Chaudhari,    M.    I.; Vanegas,    J.    M.; Pratt, L.; Muralidharan, A.; Rempe, S. B. Hydration Mimicry by Membrane Ion Channels. \textit{Ann. Rev. Phys. Chem.} \textbf{2020}, 71, 461–484. This paper\cite{ARPC} reviews local hydration structures of mono- and di-valent cations and assesses the concept of hydration mimicry for rapid transport of specific ions through ion channels. Cluster QCT and surface potentials that provide hydration free energies are also reviewed. 
 \end{itemize}
 
\section{Introduction}
This \textit{Account} describes recent research at the intersection of
the topics of ion-water
clusters,\cite{Castleman:1986fu,keesee1986thermochemical} theory of
solutions,\cite{pratt1999quasi,Asthagiri:2010tj,APP:jpcb21perspective} specific ion
effects,\cite{kunz2010specific,Zhang:2010gr,%
Pollard:2016ei} and selectivity of membrane ion
channels.\cite{ARPC} We focus on anions in water because of their
central position in classic specific ion effects, so-called Hofmeister effects.\cite{Zhang:2010gr} 
The anions considered here have
been challenges for the molecular quasi-chemical theory
(QCT),\cite{pratt1999quasi,Asthagiri:2010tj,Rogers,APP:jpcb21perspective} which is the most advanced theory available to address hydration with account of chemical-level interactions, and because  recent theoretical progress on those challenges seems
decisive.\cite{Chaudhari2017:F,ajay:F,ajay_cl,%
gomez2021free} We include HO$^-$(aq) in this discussion because of its centrality in aqueous solution chemistry, the 
continued theory and simulation interest in this ion, \cite{Asthagiri7229,marx2010aqueous,Agmon2016}, and because the new results sharpen our understanding of the hydration of 
that ion.

\subsection{Context: selectivity of membrane ion channels}
As active components of nearly half of all proteins,
\cite{Glusker:1999,Gray} ions can bind to proteins and stabilize
conformational states required for biological function, and can
participate in enzyme catalysis.\cite{mysteries}  As an example,
K$^+$ 
ions bind to membrane channel proteins and stabilize functional
conformations, thereby catalyzing permeation of K$^+$ across
cellular membranes, while also rejecting other
ions like Na$^+$.\cite{hille} 
 
Selective ion transport plays an important role in numerous
physiological functions, including electrical signaling and cell volume
control. Loss of ion selectivity, or blocking of ion transport, can have either catastrophic or beneficial effects. For example, loss of selective
conduction of K$^+$ over Na$^+$ by potassium channels in cardiac
muscle interferes with the termination of action potentials, which can
lead to life-threatening heart arrhythmia.\cite{hille} In beneficial
cases of blocked ion transport, drugs that block specific channels hold
promise for treating neurological disorders, autoimmune diseases, and
cancers.\cite{Wulff,Pardo} Peptide toxins from several poisonous
animals exemplify detrimental possibilities of blocked ion
transport.\cite{Rosenbaum} Indeed, simple divalent metal ions
can be potent channel blockers by getting trapped in the channel, and
both monovalent and divalent ions permeate selectively. Thus,
understanding the mechanisms of specific ion binding and transport in
proteins is important for understanding protein function critical to
health, disease, and therapeutic development.\cite{mysteries}

For anions such as Cl$^-$, regulation of ion concentration is achieved
through membrane transport proteins of the CLC family  and others, including channelrhodopsins,
which facilitate the passage of Cl$^-$
through electro-chemical potential gradients.\cite{alvarez2009physiology,Kim:2016,vanGordon:2021}
While  those structurally diverse proteins select for Cl$^-$, other channel proteins
discriminate against Cl$^-$. An interesting example is FLUC, a family of
fluoride-specific ion channels with dual-topology
architecture.\cite{stockbridge2015crystal} These channels display an
astonishingly high selectivity of $10^{4}$ for F$^{-}$ over Cl$^-$
despite their identical charge and their size similarity. 
Understanding such mechanisms for selectivity in ion transport has been
a target for many modeling and simulation studies
of high variety, emphasizing K$^+$/Na$^+$ selectivity of potassium ion
channels; for example, see Refs.\ \citenum{varma2007tuning,
varma2008valinomycin,Fowler2008,
Varma:2008jacs,Varma:2010,furini2011,
rossi,medovoy:2016kr,deGroot2018,jing2021thermodynamics}.  
But computational studies of anion transport mechanisms  
demonstrated by  chloride-selective channels\cite{ko2010chloride,kuang2008transpath,yin2004ion,%
chen2016free} that address comparison to 
alternatives such as FLUC are less mature.\cite{ajay_cl,Voth:2022}

A conceptually natural strategy to address ion selectivity with
computation would be direct  simulations controlling for the contrasting
cases, or perhaps a clear, quantitative theory that permits controlling
for mechanistic features of the ion binding or transport. Both of these
requests are difficult.  Here, we work toward the second of these
alternatives by building statistical molecular theory of ion binding
thermodynamics. 

\begin{figure*}[ht]
\includegraphics[width=0.9\textwidth]{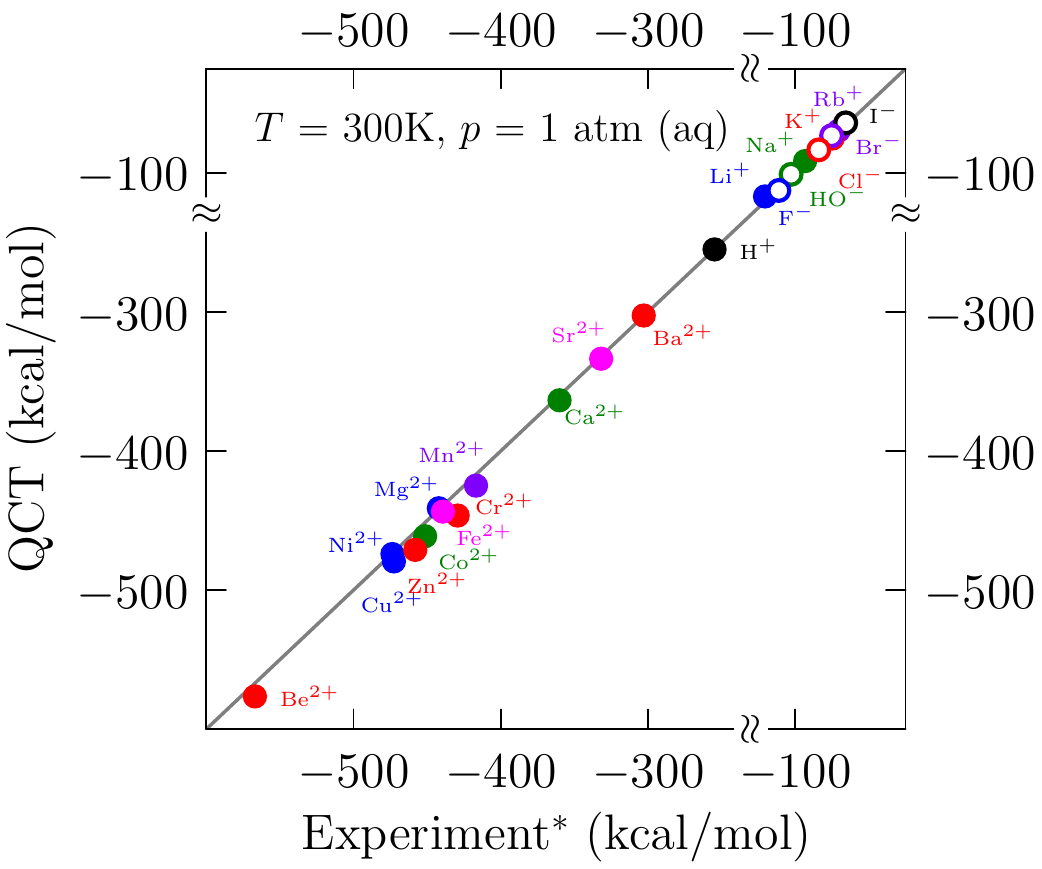}
\caption{QCT hydration free energies,
$\mu^{\mathrm{(ex)}}_{\mathrm{X}}$, for several aqueous ions. Values for
anions are shown in open circles. The computed value for
Ni$^{2+}$(aq) is taken from the reevaluation of
Ref.~\citenum{jiao2011first}. Results for H$^{+}$(aq) are from
Refs.~\citenum{ymarc91} and~\citenum{asthagiri2003jcp}.
The following discussion unpacks the
QCT theory applied. These results are the most deliberate attempt at
\textit{ab initio} evaluation of these free energies.  The values shown
are all much larger in magnitude than $k_{\mathrm{B}}T$. }
\label{fig:Exp_QCT_2021} 
\end{figure*}

\subsection{Free energies of binding of hydrated ions span a 
\textit{chemical} scale of energies }
QCT  \cite{pratt1999quasi,Asthagiri:2010tj,Rogers:2011}
aims to evaluate interaction free energies of ions in 
solution\bibnote{The asterisk on \textit{Experiment$^\ast$} of Figure
\ref{fig:Exp_QCT_2021} emphasizes that these values are not measureable
thermodynamically, but are inferred from thermodynamic experiments
together with extra-thermodynamic assumptions. In addition, as noted
previously, the  Marcus tabulation used
here\cite{ymarc91}  identifies a required standard state adjustment of \emph{in}correct sign;  See Eq. (4.6) of
Ref.~\protect\citenum{marcus2015ions}. }   
(Figure~\ref{fig:Exp_QCT_2021}) and protein binding
sites.\cite{varma2007tuning,varma2008valinomycin,Varma:2008jacs,Varma:2010,varma2011design,Stevens:2016,asthagiri2004hydration,Chaudhari:2014wb,Asthagiri:2010tj,%
Rogers:2011,asthagiri2003jcp,%
Jiao:2012,Dudev2013,rossi,chaudhari2018SrBa,ARPC} That
interaction free energy, or excess chemical potential, 
\begin{eqnarray}
\mu_{\mathrm{X}}^{(\mathrm{ex})} = \mu_{\mathrm{X}}  - k_{\mathrm{B}}T \ln \rho_{\mathrm{X}} V/\mathcal{Q}\left\lbrack \mathrm{X}\right\rbrack
\label{eq:ideal}
\end{eqnarray}
is obtained from the full chemical potential $\mu_{\mathrm{X}}$ 
less the indicated ideal
contribution. Here $T$ is the temperature, $k_{\mathrm{B}}$ Boltzmann's constant, $\rho_{\mathrm{X}}$ the
number density of the ion of interest,  and
$\mathcal{Q}\left\lbrack\mathrm{X}\right\rbrack$ is the canonical
partition function of a   molecule X in volume $V$.\cite{lrp:book,ARPC} 

These free energies --- single-ion activities when X is an ion --- are
knowable and appropriate targets for computation.  But they are not
measured on the basis solely of classic thermodynamics.  Thus,
widely available tabulations --- Figure~\ref{fig:Exp_QCT_2021} utilizes
one such tabulation --- adopt extra-thermodynamic assumptions.

Observe (Figure~\ref{fig:Exp_QCT_2021}) that the free
energies span a \textit{\underline{chemical} energy} scale, much larger
than $k_{\mathrm{B}}T \approx 0.6$~kcal/mol at room temperature. 
For example, 
the hydration free energy of Be$^{2+}$ is about $1000\, k_{\mathrm{B}}T$. 
Clustered below
$-400$~kcal/mol are values for divalent transition metal
ions.\cite{asthagiri2004hydration,jiao2011first}
Including the
aqueous ferric ion, Fe$^{3+}$(aq), for which 
QCT performs satisfactorily,\cite{10.1021/jp980229p} would require expanding the
range of Figure~\ref{fig:Exp_QCT_2021} by another factor 2.  Thus, though
the now-canonical van der
Waals perspective on
liquids is an appropriate definition of the statistical mechanical
problem for treating liquids generically,  it 
immediately emphasizes $<$1\%-magnitude
free energy effects. In contrast, the basic concept of QCT is to treat an ion together with
inner-shell partners as an individual molecular 
species.\cite{10.1021/jp980229p,Rempe:2000uw,%
pratt1999quasi,pratt2007potential,%
Asthagiri:2010tj,Rogers}  
Interactions of typical ions with inner-shell partners  are chemical in
nature, molecularly intricate, and intense on a thermal scale.   The concept
of ``inner-shell'' partners is central to QCT. It identifies near-neighbors of a 
targeted species, and will be discussed 
later
for the present applications.

Molecular quasi-chemical theory
(QCT)\cite{pratt1999quasi,pratt2007potential,%
Asthagiri:2010tj,APP:jpcb21perspective} developed with the explicit goal of including chemical level interactions within a molecular statistical thermodynamic theory. Initial applications were simple and
remarkably accurate.\cite{Rempe:2000uw} That success obviates a canonical `molecular force-field fitting $\rightarrow$ molecular simulation' workflow in the study of liquids. 

\subsubsection{Direct QCT}
The basic status of QCT may be supported by the fact that QCT itself can
be implemented through molecular simulation
calculations.\cite{APP:jpcb21perspective} That approach is termed
\textit{`direct'} QCT.\cite{Sabo:h2} In the direct approach, we acknowledge and exploit
the spatial dependence of solute-solvent interaction strengths.
The short-range interactions can be chemically involved,
and for a suitable choice of the inner-shell, the
long-range interactions admit a Gaussian statistical 
model. This direct QCT approach parses the 
hydration free energy into physically meaningful and computationally
well-defined chemical, packing, and long-range non-specific
contributions, thereby becoming a framework to
conceptualize molecular solutions.\cite{Rogers} 

Direct QCT works naturally with common simulation packages based on
either empirical classical or \textit{ab initio}
forcefields.\cite{APP:jpcb21perspective} On that simulation basis, QCT
provides a compelling molecular theory of liquid water
itself.\cite{Shah:2007dm,Weber:2011hd,chempath2009quasichemical} The direct QCT approach enabled
the first direct calculation of the hydration free energy of a
protein.\cite{Weber:2012kc} Subsequent studies have highlighted the
limitations of additive models of free energies that are \textit{de
rigueur} in biophysical speculation, and recently led to
transformative insights into decades-old assumptions about hydrophobic
hydration in proteins.\cite{tomar:jpcl20} 
   
Nevertheless, the initial motivation was the exploitation of molecular
electronic structure calculations within statistical thermodynamic
modeling.\cite{10.1021/jp980229p} That approach is called
\textit{`cluster'} QCT, wherein the chemical contribution noted above is
related to \textit{physical} solute-solvent clusters. The connection
between direct and cluster approaches has been deliberately discussed
elsewhere.\cite{Asthagiri:2010tj,APP:jpcb21perspective,Rogers,Chaudhari:2017gsa}

\subsubsection{Cluster QCT}

Cluster QCT provides  a concise format, 
\begin{multline}
\mu^{\mathrm{(ex)}}_{\mathrm{X}} = -k_{\mathrm{B}}T\ln K^{(0)}_{n}\rho_{\mathrm{H_2O}}{}^{n}  \\
+ k_{\mathrm{B}}T\ln p_{\mathrm{X}}(n) \\
	+\left(\mu^{\mathrm{(ex)}}_{\ce{(H2O)_nX}}-n\mu^{\mathrm{(ex)}}_{\mathrm{H_2O}}\right)~,
		\label{eq:qctX}
\end{multline}
for the free energies that we seek.  This is 
exact statistical thermodynamics,\cite{Asthagiri:2010tj}
and provides a foundation for mixed resolution approaches, such as QM/MM. We will discuss these three terms
in turn.  The first term on the right of Eq.~\eqref{eq:qctX} is the
\textit{inner-shell} contribution.   It is obtained by 
studying $n$ water molecules clustering with the X species of
interest,\cite{gomez2021free} but without the exterior solution. 
This contribution involves the equilibrium 
constant, $K^{(0)}_{n}$, discussed
in Sec.~\ref{sec:innershell} below. By
utilizing the solution density of the ligands, $\rho_{\mathrm{H_2O}}$,
this term properly  assesses the availability of the water molecule
ligands.

The rightmost term of Eq.~\eqref{eq:qctX} is the \textit{outer-shell}
contribution.   This term involves the hydration of the identified
\ce{(H2O)_nX}  cluster, addressing interactions of the cluster
with the exterior solution
environment.  Previous applications of QCT have
utilized the polarizable continuum model (PCM)\cite{Tomasi:2005tc}
for this task (see Refs. \citenum{rempe2004inner} \& \citenum{Sabo:2013gs} as examples).
PCM has been incorporated into standard electronic structure codes
and focuses on the interactions with the
solution at long-range.  Nevertheless, it is approximate at a molecular
scale, and a symptom of that approximate character is the sensitivity of
PCM results to radii-parameters that are required. We will  
discuss that issue further below.

The remaining term of Eq.~\eqref{eq:qctX} involves the probability
$p_{\mathrm{X}}(n)$ that $n$ water molecules contact a
distinguished X during its physical motion in solution. This
term describes the polydispersity of the populations of an X
inner-shell;  if only one size, say $n$, were possible,
then  $\ln p_{\mathrm{X}}(n)=0$. Determination of $p_{\mathrm{X}}(n)$
requires adoption of a proximity criterion describing
how a water molecule contacts an X.\cite{gomez2021rough,Chaudhari2017:F,ajay:F,ajay_cl} 
That polydispersity
contribution is typically the smallest of these three,
is conceptually simplest, and here we utilize AIMD
simulation of the solution of interest to compute this term.

\subsubsection{Anion Hydration}
\begin{figure}[!htbp]
\centering 
\includegraphics[width=0.9\linewidth,height=0.9\linewidth]{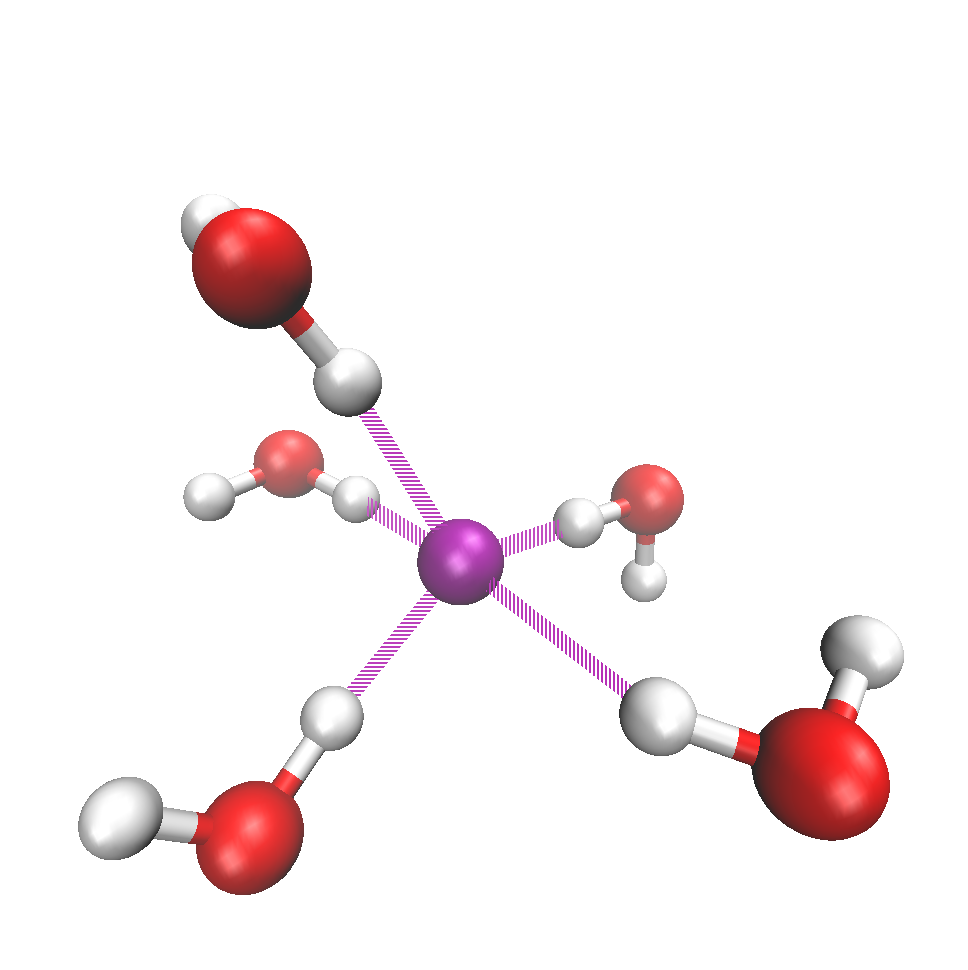}
\caption[inner]{A structure sampled
from the AIMD trajectory for the isolated \ce{(H2O)5F}$^-$  cluster.  
}
\label{fig:Outline}
\end{figure}

QCT 
applies to both cation and anion hydration cases. In contrast to
cations, however, anion hydration clusters often exhibit H-bond donation to the
ion (Figure~\ref{fig:Outline}; see also Ref.~\citenum{gomez2021rough}).  Anion-hydration
clusters can be structurally
delicate, specifically involving ligand-ligand hydrogen bonds, and that can make hydrated anions  
more challenging cases.

\begin{figure}[ht]
\includegraphics[width=0.4\textwidth]{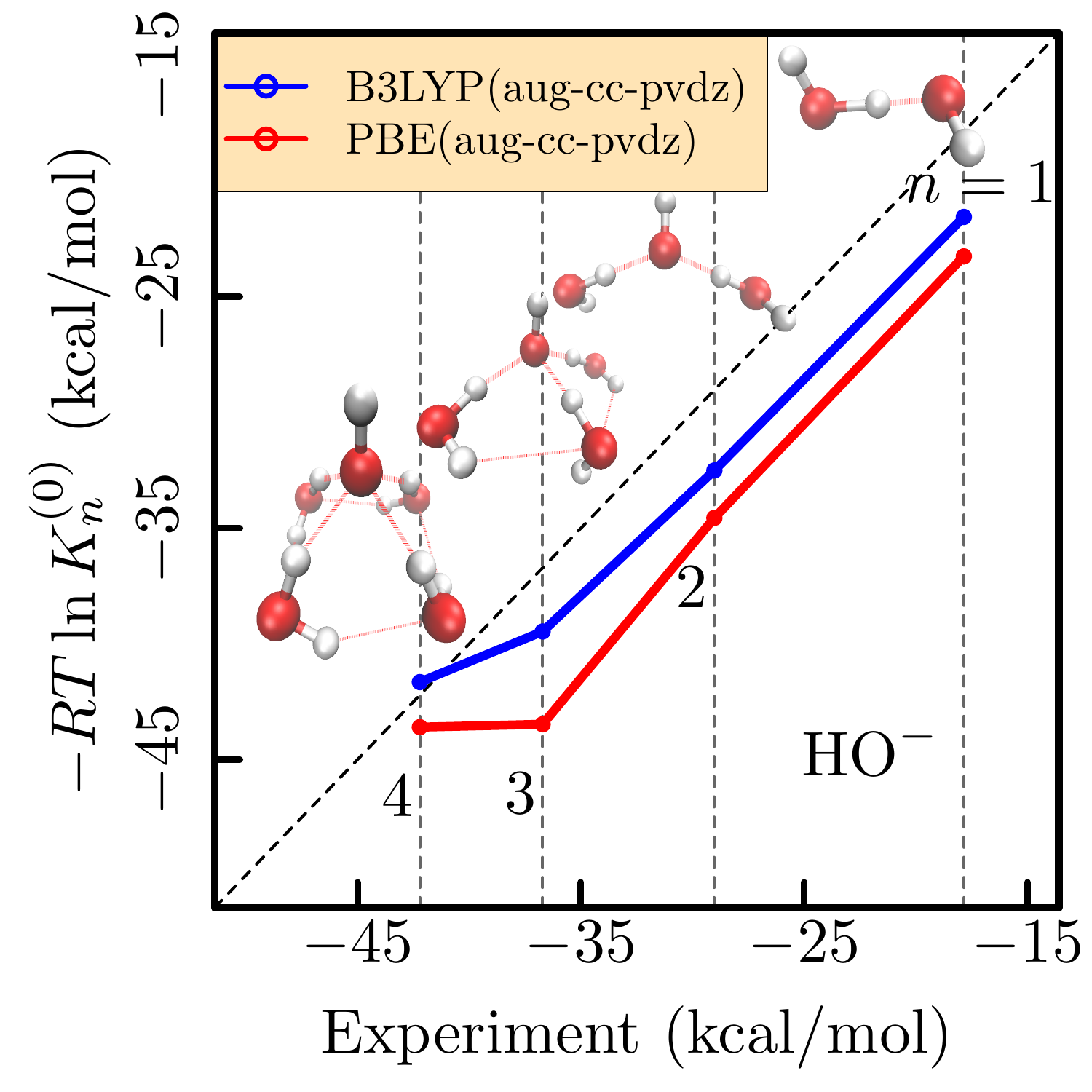} 
\caption{Cluster free energies using harmonic approximation compared
with experimental values 
for $(\mathrm{H}_2 \mathrm{O})_n\mathrm{HO}^-$,
and $1\leq n \leq 4$. 
The  implicit  density  is $\rho_0=p/k_{\mathrm{B}}T$ with $p=  1$~ atm
and $T= 300$~K. 
The molecular graphic insets show optimized structures for
each $n$.
The legend
indicates electron density functional and basis sets, 
thus demonstrating the sensitivity to those aspects of 
these calculations.  $K^{(0)}_{n}$ is the traditional
equilibrium constant, introduced 
with Eq.~\eqref{eq:2charged}  below. } 
\label{fig:HO_harmonic} 
\end{figure}

Initial QCT applications  to hydrated anions worked 
simply with reasonable
accuracy\cite{asthagiri2003jcp,Chaudhari2017:F,ajay:F,ajay_cl} compared with experiments.\cite{tissandier1998proton}
Nevertheless, specifics of the technical ingredients can be
perplexing (Figure~\ref{fig:HO_harmonic}). Refinement of those initial 
applications has led to the further
considerations  discussed here;
specifically, treatment of anharmonic effects on free energies of
 ion hydration clusters, and the status of the polarizable continuum
model\cite{Tomasi:2005tc} (PCM) for the hydration free energy of those
clusters.

\subsection{Theory implemented with simulation data}

We implement QCT here by bringing together quantities available
from several different standard computations. The simulation work here
thus contrasts with
\textit{``let's take a look''}  direct numerical simulations.
Indeed, QCT seeks to define minimal clusters that 
provide the  information necessary for the statistical thermodynamic
theory;  thus, we explicitly do not attempt to 
construct observational dynamical simulations, nor
do we seek a large cluster-size limit in our simulation calculations.\cite{basdogan}
Before returning to discuss the theory, we note the  
details required for the computational procedures in the following.
In addition, AIMD simulations of these  systems are
indeed readily available, and those observational 
calculation help secure details that fill-out our understanding of
these systems.  Here, we note
some of that previous
work.

The extended work of Heuft and 
Meijer\cite{Heuft:2003iva,Heuft:2005jt,Heuft:2005kx}
initiated AIMD calculations on halide
anions in water.  They noted that residence
times of water ligands in a halide 
inner shell 
spanned approximately 8~ps, 12~ps, and 17~ps
for I$^-$, Cl$^-$, and F$^-$, respectively.  Those time scales 
are readily accessible by current AIMD calculations.

The interesting work of Wiktor, \textit{et al.,}\cite{wiktor2017partial} focused on a
basic thermodynamic quantity, the partial 
molar volumes of ions in water.  Such studies are likely to provide 
fruitful next steps in the understanding of these systems.

The AIMD of Duignan and co-workers\cite{Duignan:2017iha,Duignan2021} on F$^-$(aq)
also focused on a basic thermodynamic characteristic of 
simple ions in water, namely hydration free energies.
They made a case for application of ultra-high accuracy 
electronic structure calculations to these problems.
Our discussion below will identify aspects of 
the present efforts that overlap with that previous work, but support a different conclusion:
 specifically, standard electronic structure calculations, properly 
integrated into statistical thermodynamic theory, are sufficient for 
experimental accuracy.

Finally, for this section, we note the
extensive AIMD work on HO$^-$(aq) that
has been exhaustively reviewed.\cite{marx2010aqueous,Agmon2016}
Though that work did not proceed to evaluation
of standard thermodynamic characteristics, the discussions below will elaborate on
specific points of comparison.

\subsubsection{Procedures for bulk X(aq) solutions}\label{subsecMethodsbulk}
The data utilized here for F$^-$(aq) and Cl$^-$(aq) was obtained 
from previous work\cite{ajay:F,ajay_cl} that treated a single ion and 64 water molecules
using the VASP simulation
package.\cite{kresse1996efficient} The system was a cubic cell of 
edge 1.24~nm with  periodic boundary conditions. The PW91
generalized gradient approximation described the core-valence
interactions using the projector augmented-wave (PAW) method. Plane
waves with a kinetic energy cutoff of 400eV and a time step of
0.5~fs were used for the simulation in the NVE ensemble. A temperature of
350~K was targeted for the  simulation to avoid glassy behavior that
can result at lower $T$.\cite{ajay:F,ajay_cl}
After discarding 50~ps of trajectory as aging,
our analysis was based on a
50~ps production trajectory.

\begin{figure*}[!htbp]
\centering 
\includegraphics[width=0.3\linewidth]{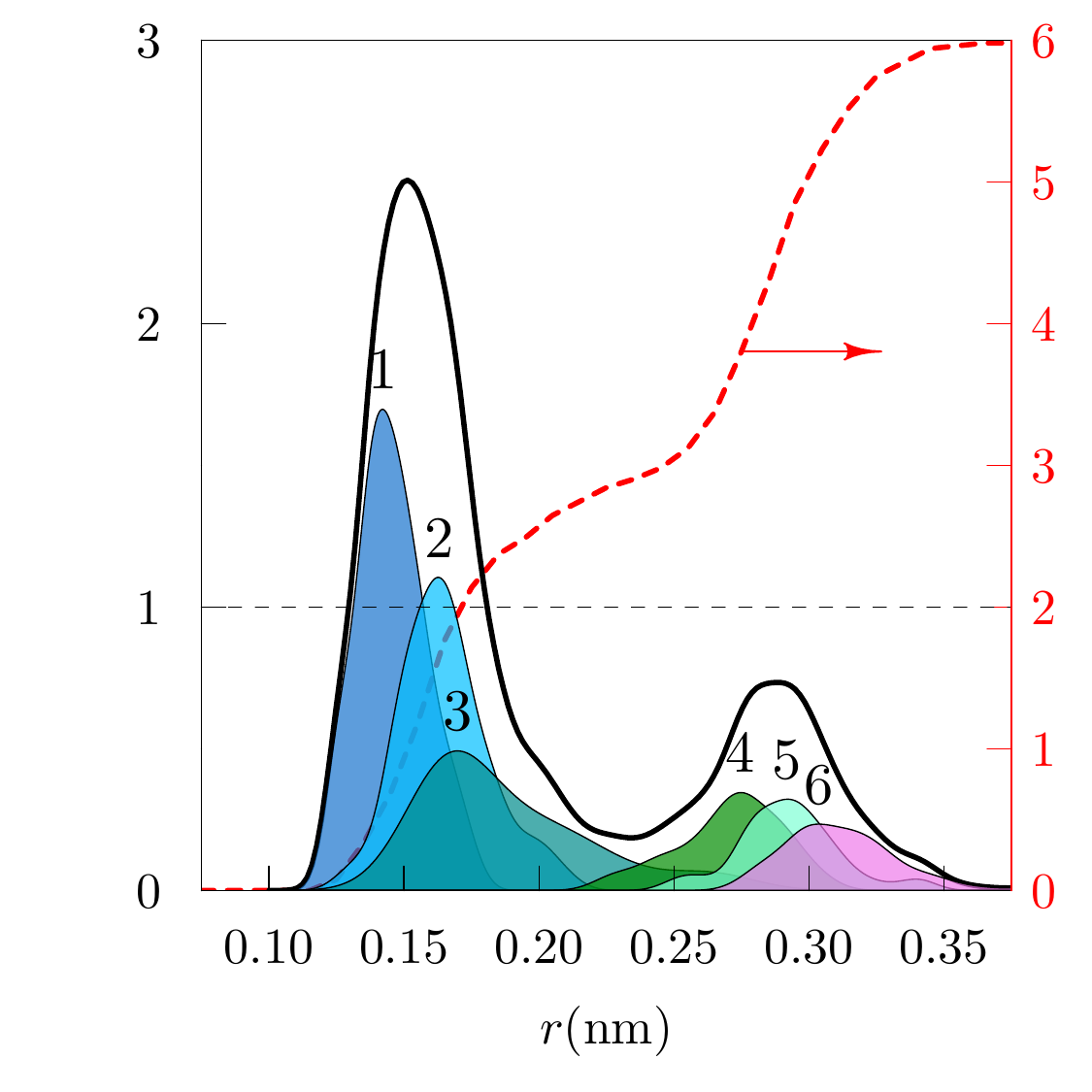} \hfil
\includegraphics[width=0.3\linewidth]{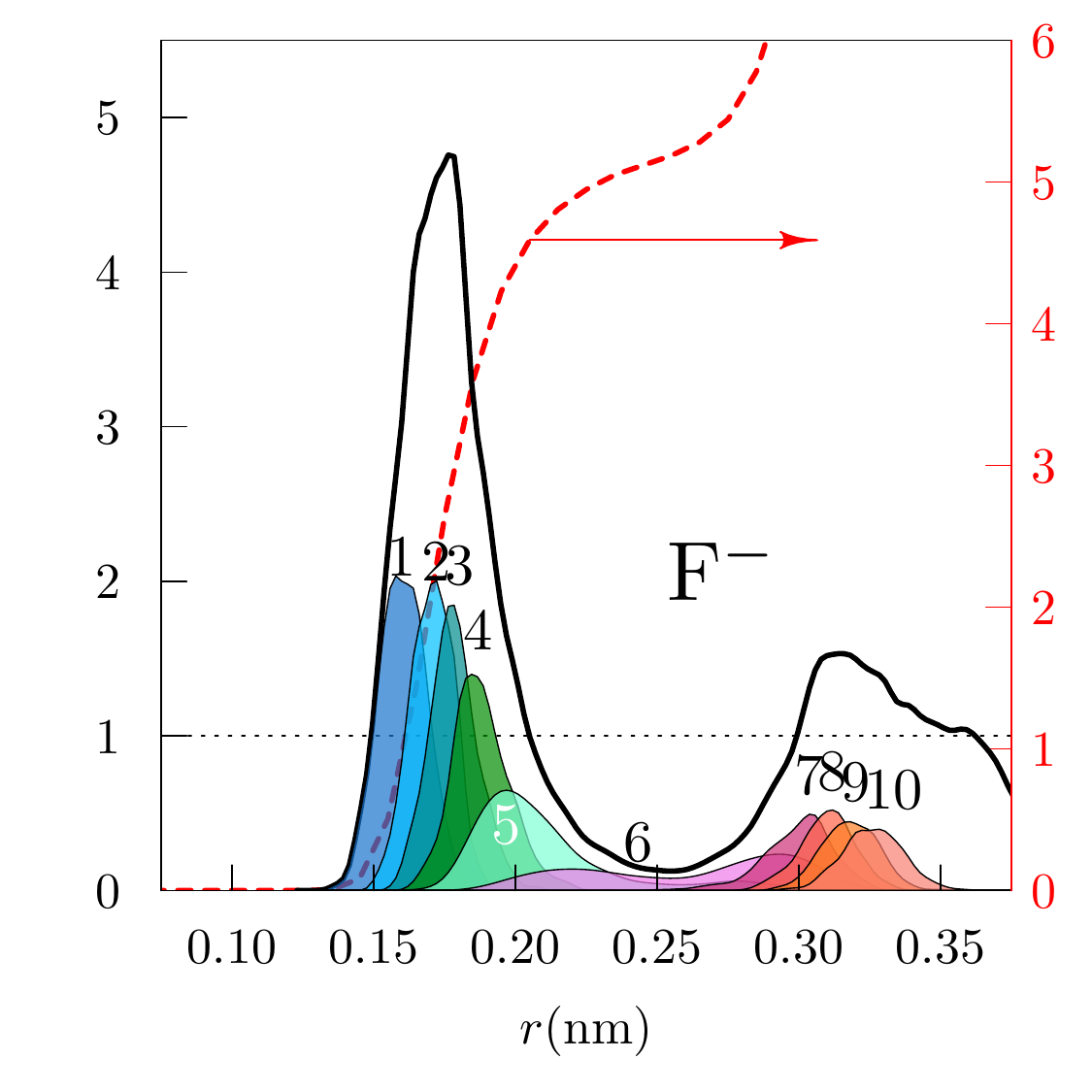} \linebreak
\includegraphics[width=0.3\linewidth]{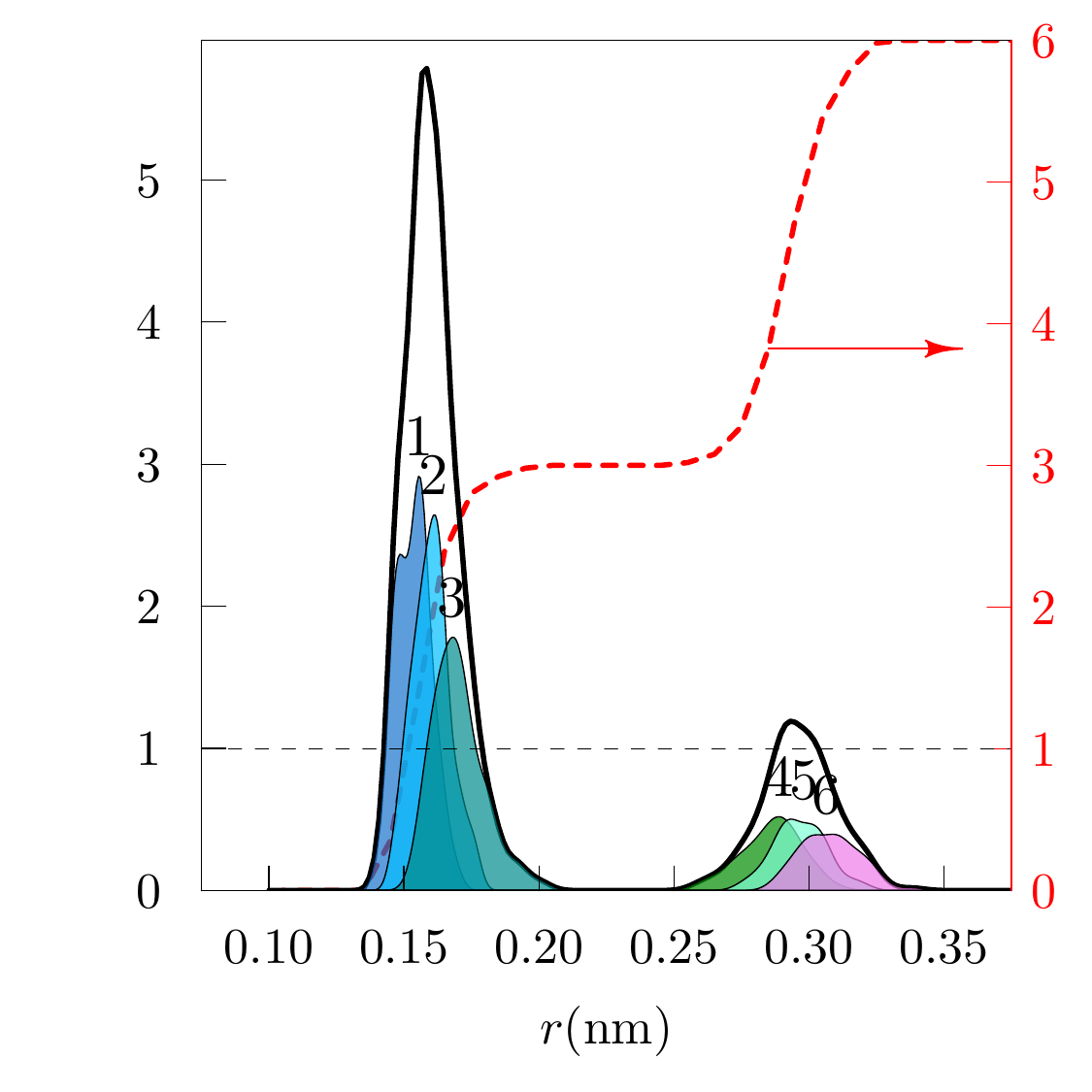}\hfil
\includegraphics[width=0.3\linewidth]{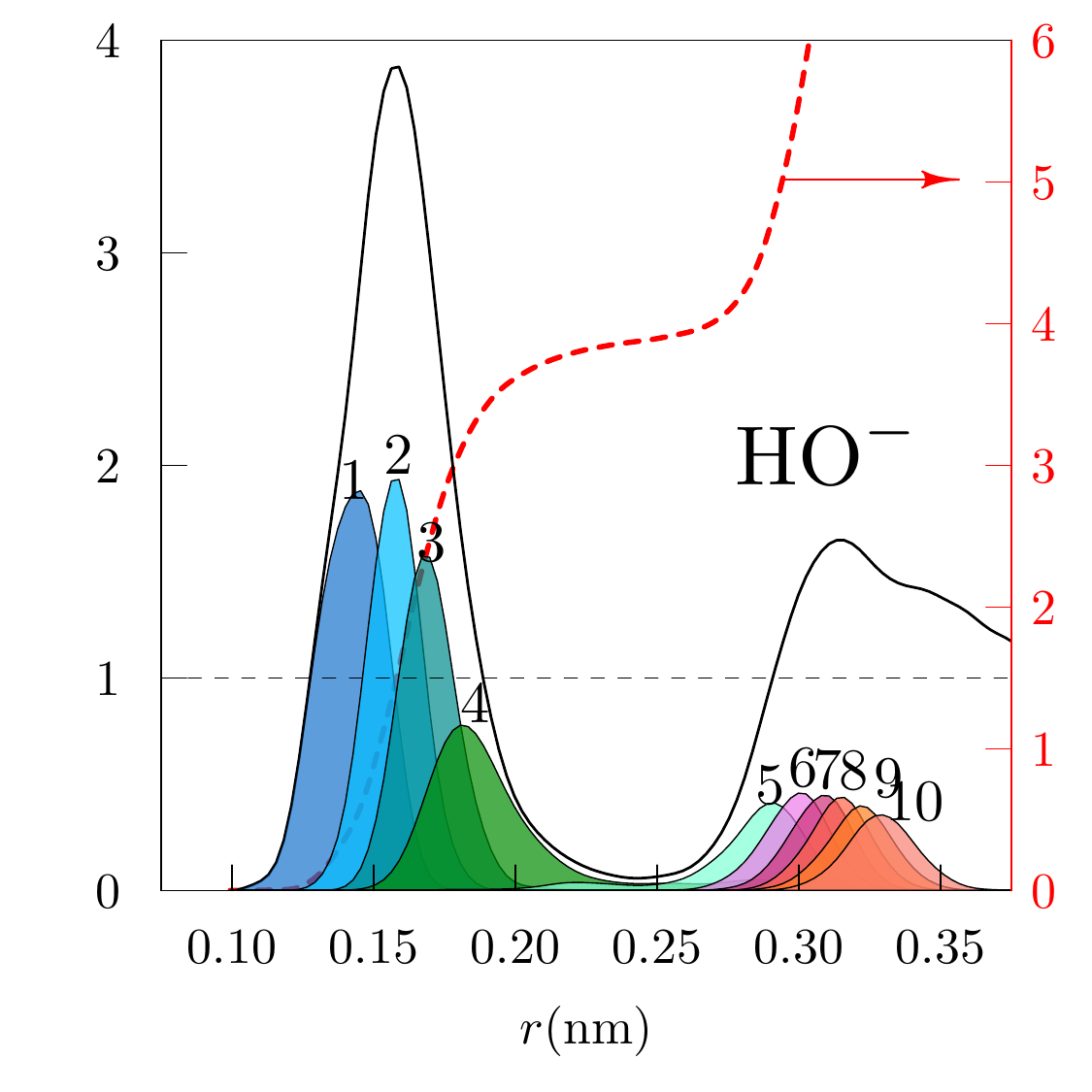} \linebreak
\includegraphics[width=0.3\linewidth]{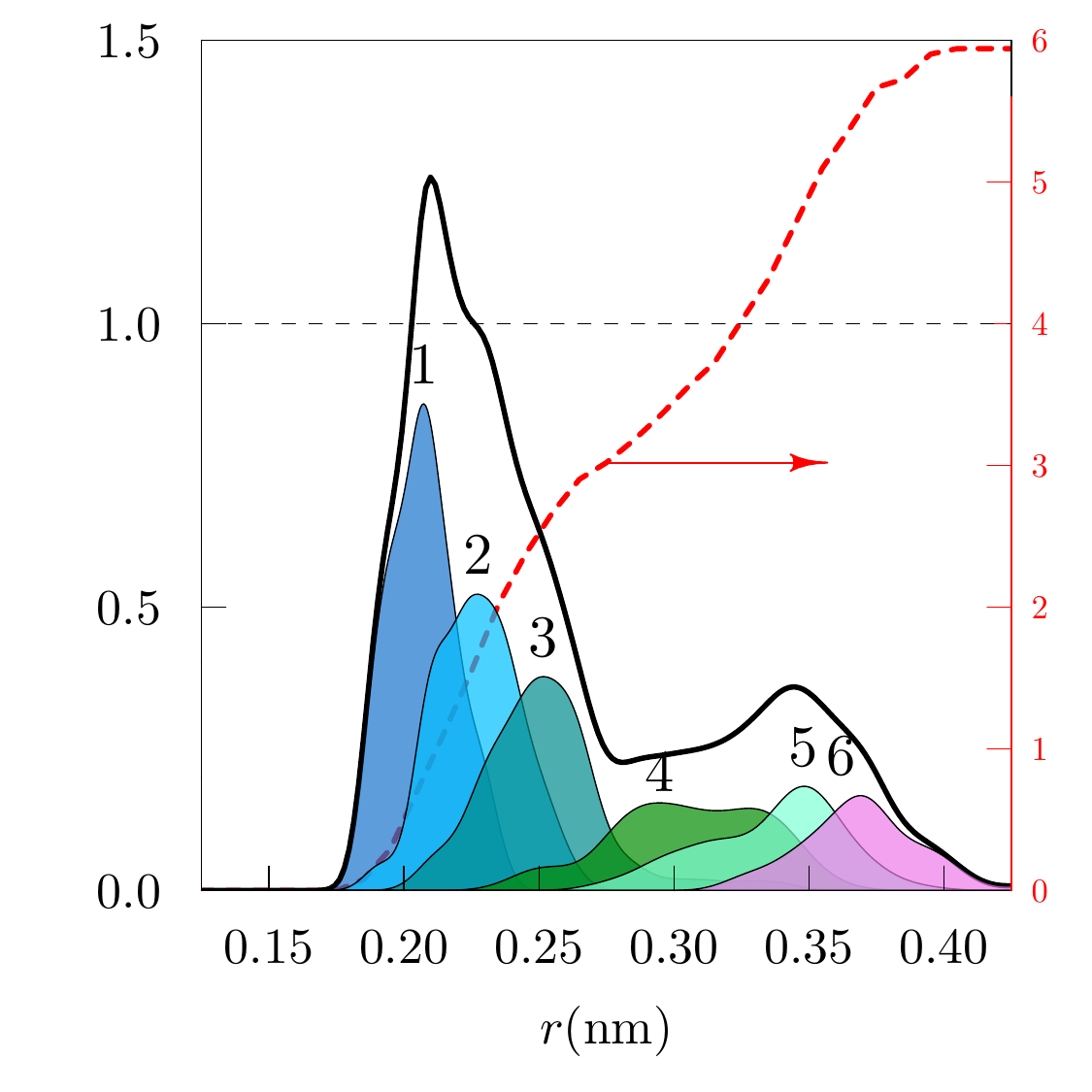} \hfil
\includegraphics[width=0.3\linewidth]{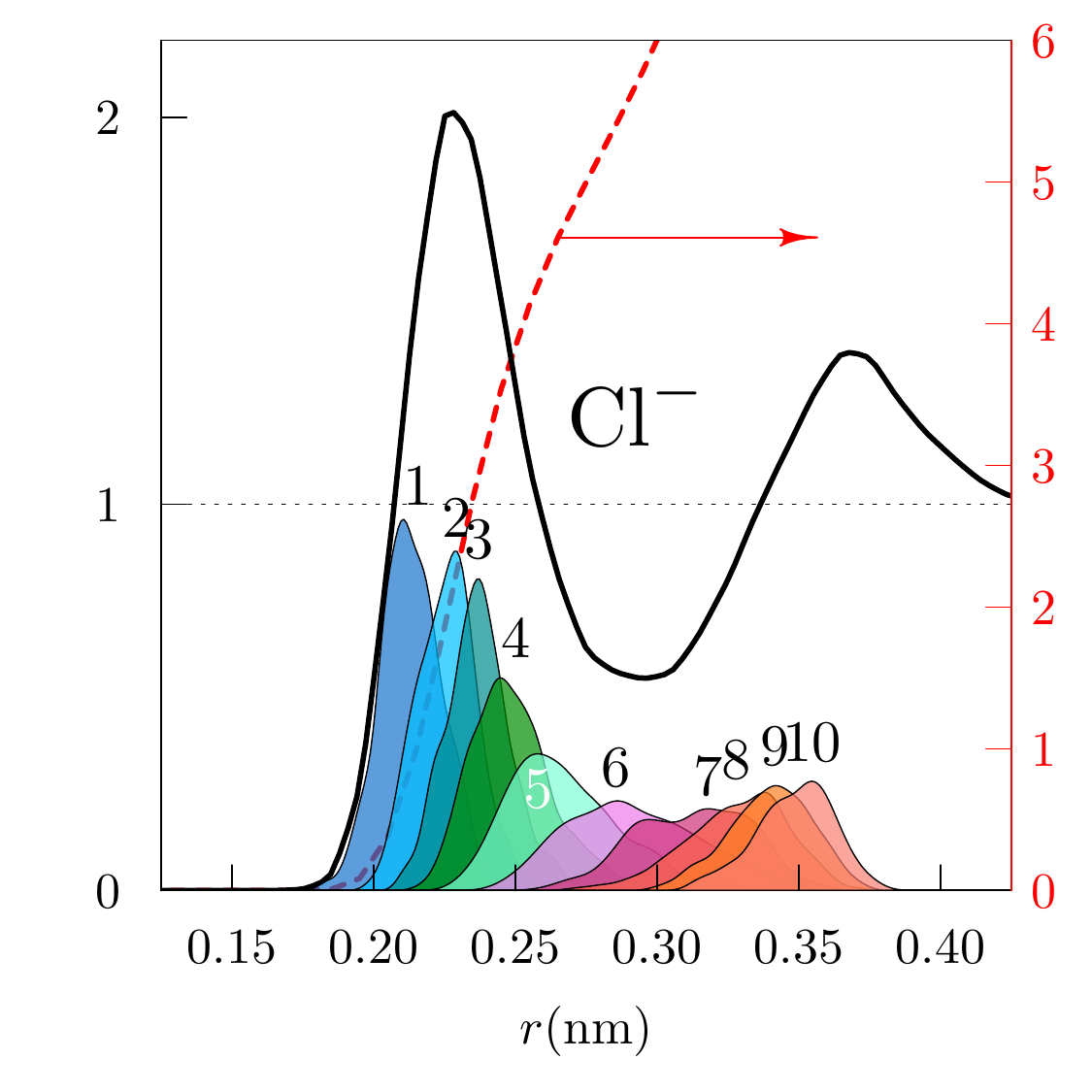}
\caption[HOcompare]{Radial distributions of \ce{H2O} molecule H atoms
relative to the  X anion from AIMD trajectory, left panel for the
isolated \ce{(H2O)3X} cluster, right panel for 
the X(aq) with periodic boundary conditions 
corresponding to our
thermodynamic state.  The neighborship-ordered
distributions on the left panel are normalized 
to
$1/\varrho_{\mathrm{H}},$ with $\varrho_{\mathrm{H}}$ the number density
of solvent H-atoms for the right panel.   These distributions are thus
directly comparable. The red dashed curves (right-side axes) give the
running H-atom coordination number, $n_{\mathrm{H|X}}(r)$. For the
\ce{(H2O)3HO}$^-$ cluster, $n_{\mathrm{H|O}}(r)$ plateaus near 3,
indicating simple H-bond donation.  For the bulk solution,
$n_{\mathrm{H|O}}(r)$ plateaus at about 3.8.}
\label{fig:HOcompare}
\end{figure*}

For HO$^-$(aq), Br$^-$(aq) and I$^-$(aq), the AIMD calculations  
are new here, and used the CP2K simulation
package\cite{10.1063/5.0007045} 
to treat a single ion and 64 water molecules in 
periodic boundary conditions.  
We adopted the PBE functional with  Goedecker, Teter and Hutter\cite{goedecker1996separable} (GTH) pseudopotentials in the GPW schemes,\cite{lippert1999gaussian} as broadly used and consistent with 
our previous cluster results.\cite{gomez2021rough}
Molecularly optimized DZVPMOLOPT-SR-GTH
basis sets were obtained from the CP2K website. Plane waves with a
kinetic energy cutoff of 400eV and a time step of 0.5~fs were used for
the simulation in the NVT ensemble. The cubic cell
with edge 1.27~nm reasonably matches the
experimental density of water at our standard conditions.
$T = 300$~K was selected\cite{NOSE} through the Nos\'e-Hoover thermostat. 
Our analysis was based on
50~ps of production trajectory after 50~ps of aging.
Figure~\ref{fig:HOcompare} provides a standard overview of the
bulk solution structures observed.

\begin{figure*}
\begin{center}
    \includegraphics[width=0.45\textwidth]{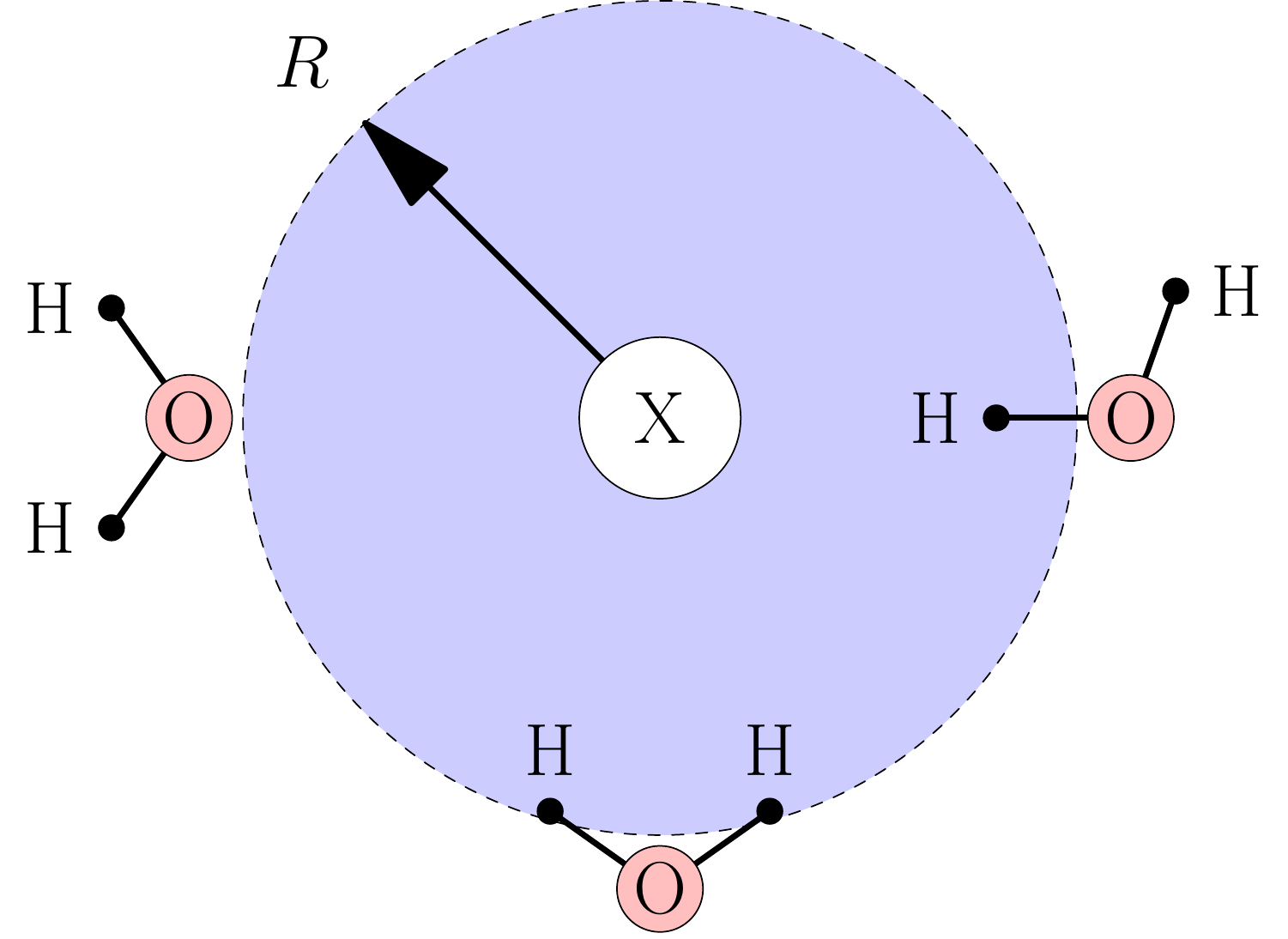} \\
    \includegraphics[width=0.45\textwidth]{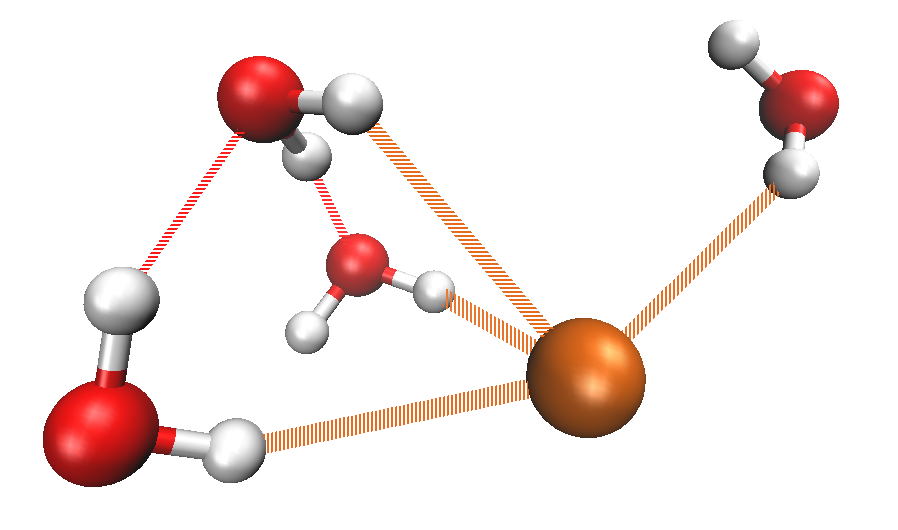} \\
    \includegraphics[width=0.45\textwidth]{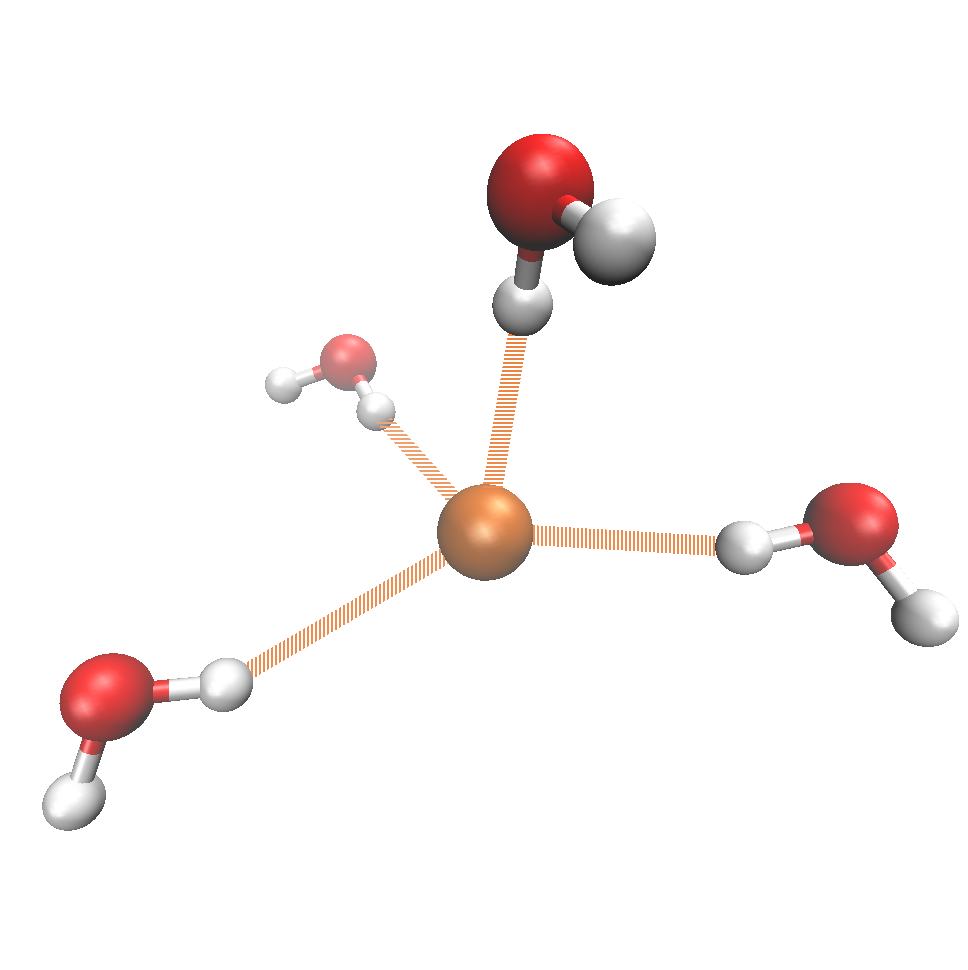}
\end{center}
\caption{Clustering of inner-shell water molecules with  central anion.  (Top panel):  The primitive clustering criterion
requiring a
clustered water molecule to donate an H atom to the sphere
within $R$ of the ion.  The rightmost water molecule is a
member of this cluster, but the leftmost is not.  The bottom water
molecule donates two H atoms though counts
only one ligand.  (Middle panel):
A clustered configuration chosen arbitrarily
from the AIMD trajectory for an isolated
\ce{(H2O)4I}$^-$.  (Bottom panel):
A clustered configuration  selected arbitrarily
from the AIMD simulation of I$^-$(aq) solution.
These graphics illustrate asymmetric anion hydration that is moderated in bulk solution (bottom) compared with clusters (middle).} \label{fig:ClusteringCriterion}
\end{figure*}

\subsubsection{Procedures for isolated $\left (
\mathrm{H}_2\mathrm{O} \right )_n \mathrm{X}$}\label{subsecMethodsGP} 
Molecular dynamics trajectories of the isolated $\left (
\mathrm{H}_2\mathrm{O} \right )_n \mathrm{X}$ clusters for $2 \leq n \leq 5$ and
X = F$^-$, HO$^-$, Cl$^-$, Br$^-$, and I$^-$ were obtained using 
CP2K,
just as for the HO$^-$, Br$^-$, and I$^-$ bulk solution
calculations specified above.
The pseudopotentials, functionals, and basis sets were the same. 
These simulations set temperatures at 300~K with the Nos\'e-Hoover
thermostat, using the GPW basis with default
settings and a kinetic energy 
cutoff of 400eV.  5~ps of production 
trajectory, with a time step of 1~fs, was analyzed after 5~ps of aging.

\begin{figure}[ht]
\includegraphics[width=0.45\textwidth]{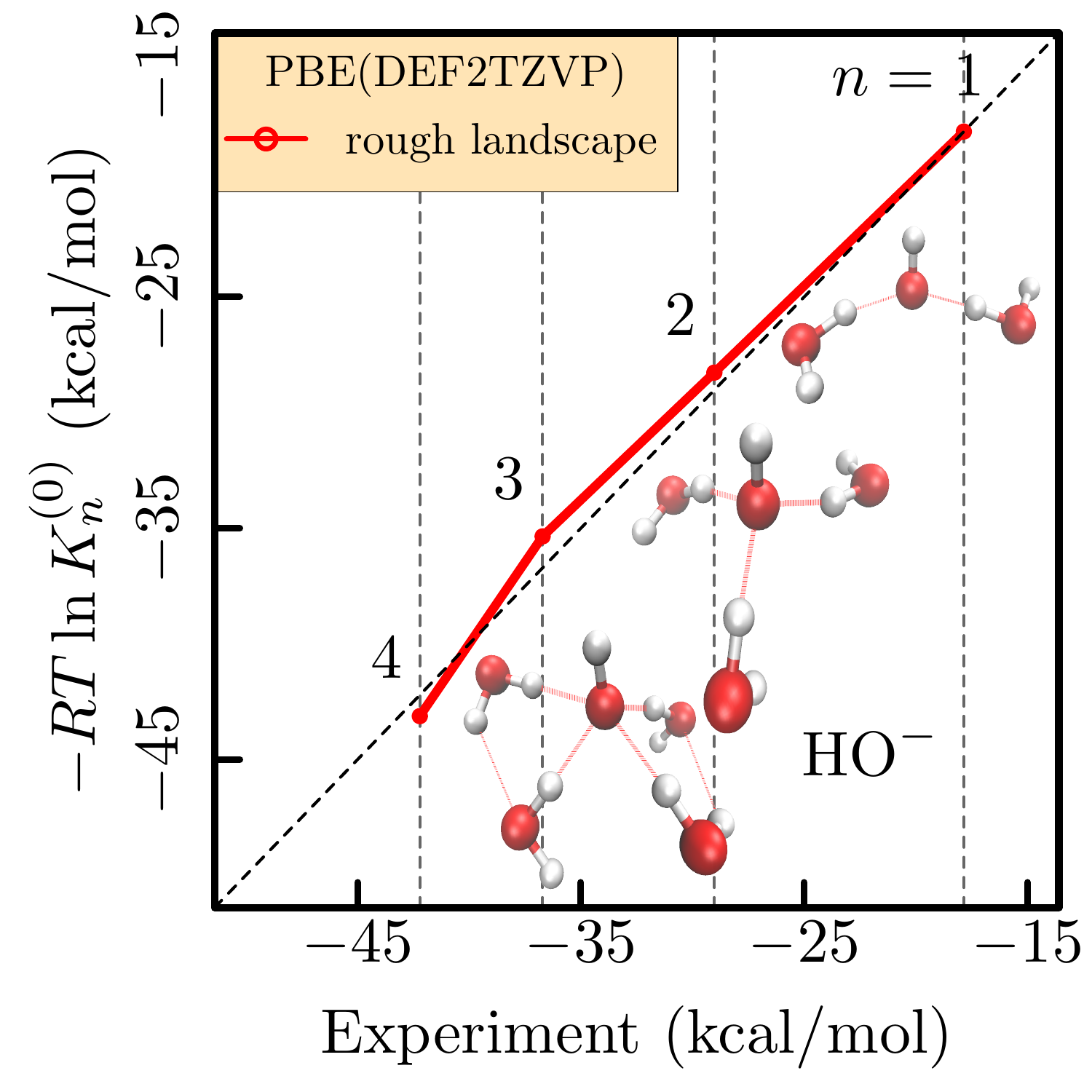} 
\caption{For HO$^-$, free energies for 
the inner-shell contributions (Eq.~\eqref{eq:qctX}).  The 
implicit density is $\rho_0 = p/k_{\mathrm{B}}T$ with
$p = 1$~atm and $T = 300$~K.  The embedded
graphics
depict structures sampled from the 
AIMD trajectory.} 
\label{fig:HO_innershell_PBE} 
\end{figure}

For our statistical, or \textit{\underline{rough landscape}}, analysis of
those trajectories, cluster structures were  screened for
consistency with our
clustering definition (Figure~\ref{fig:ClusteringCriterion}). Each
clustered configuration is analyzed,
according to the right-hand side of Eq.~\eqref{Eq:final}, and the separate structures subjected to single-point calculations
using 
the Gaussian\cite{g16} electronic structure software with the PBE functional and
DEF2TZVP basis set for F$^-$, HO$^-$, Cl$^-$, Br$^-$, and I$^-$  configurations.  Using
the resulting thermal averaging in Eq.~\eqref{Eq:final}, and $K_1^{(0)}$
from experiment, the resulting step-wise $K_n^{(0)}$ (Figure~\ref{fig:HO_innershell_PBE}) produce the
accurate results used below.\cite{gomez2021free,gomez2021rough}

\section{Discussion and Results}

\subsection{Inner-shell contributions}\label{sec:innershell}

Study of the associative equilibria
\begin{equation} n \ce{H2O}  + \mathrm{X} \rightleftharpoons 
\ce{(H2O)_n X} 
\label{eq:2charged} 
\end{equation} 
is a basic feature of QCT.
Here X
$\equiv$ F$^-$, HO$^-$, Cl$^-$, Br$^-$, or I$^-$.
Eq.~\eqref{eq:2charged} directs attention to \begin{eqnarray}
K^{(0)}_n = \frac{ \rho_{\ce{(H2O)_n X}} }{
\rho_{\ce{H2O}}{}^n \rho_{{\mathrm{X}}} }~, \label{eq:Kratio}
\end{eqnarray} where $\rho_{\ce{(H2O)_n X}}$ 
is the number
density of \ce{(H2O)_n X} species.  
 $K^{(0)}_{n}$ requires   
definition of formed \ce{(H2O)_n X} clusters for evaluation of actual 
densities. Such definitions
amount to defining proximity of a \ce{H2O} ligand to an X ion. Although
judgement might be required for a natural
proximity definition, here we defer discussion of that
definition until after subsequent QCT developments.

Our scheme for evaluating $K_{n}{}^{(0)}$ is anchored in classic 
statistical thermodynamics,\cite{pratt1999quasi,ajay_cl,lrp:book} and proceeds
incrementally following 
\begin{eqnarray}  K_{n}{}^{(0)} =
\frac{
K_{1}{}^{(0)}K_{n-1}{}^{(0)}
}{
n\left \langle \me^{\beta \Delta U_n}
\right \vert n\rangle}~. 
\label{Eq:final} 
\end{eqnarray}
This formulation introduces the energy differences 
\begin{multline} \Delta U_n =
\left\lbrace U\left\lbrack\ce{(H2O)_n X} \right\rbrack -
U\left\lbrack\ce{(H2O)_{n{-}1} X} \right\rbrack\right\rbrace \\
- \left\lbrace U\left\lbrack\ce{(H2O)X}\right\rbrack - U\left\lbrack\mathrm{X}\right\rbrack\right\rbrace~.
\label{Eq:step_fe_2} 
\end{multline}
The brackets of Eq.~\eqref{Eq:final},
$\langle \ldots \vert n\rangle$, indicate the thermal average utilizing 
the canonical simulation stream for the $\ce{(H2O)_n
X}$ cluster.\cite{ajay_cl}  

\begin{figure}[ht]
\includegraphics[width=0.4\textwidth]{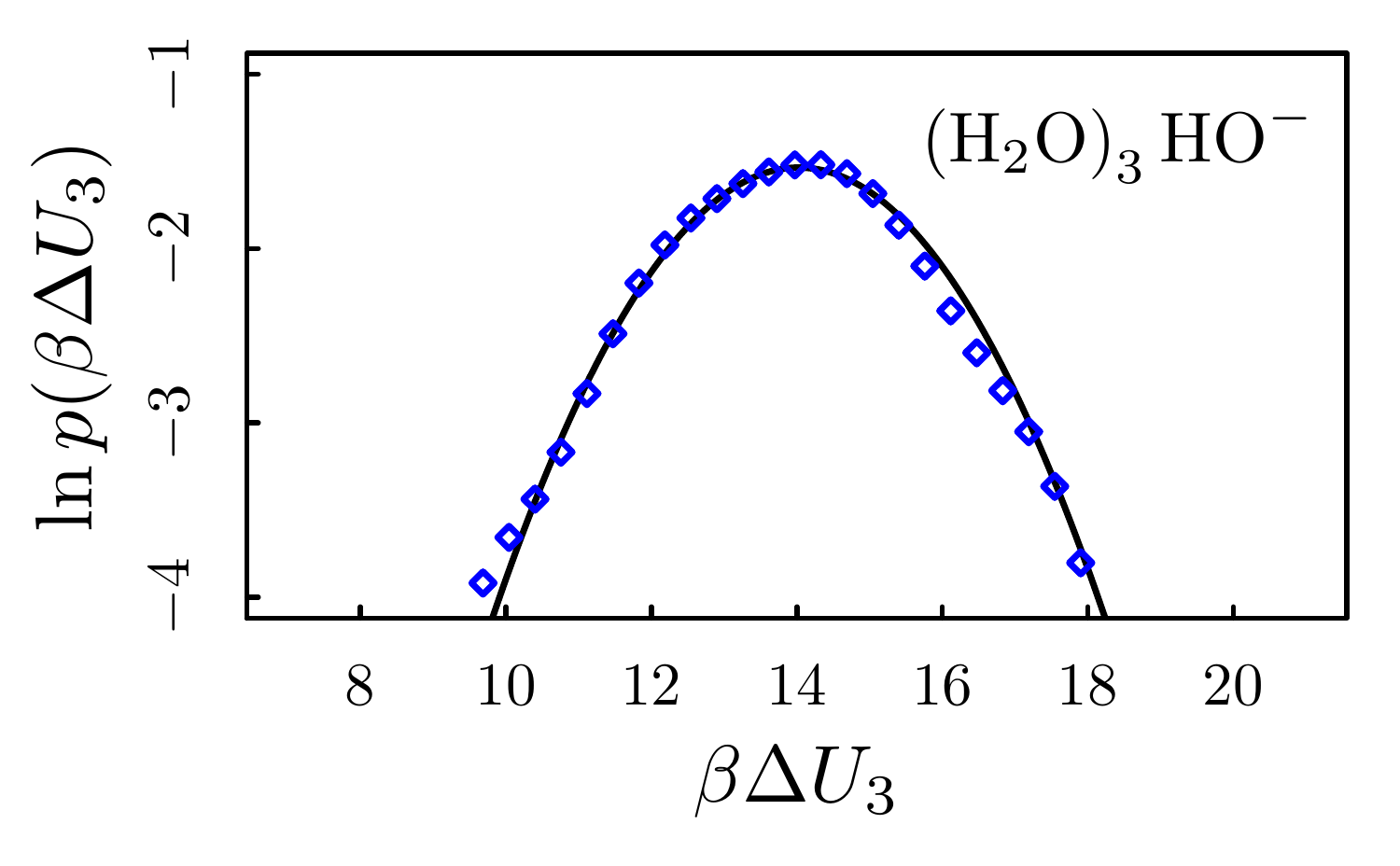}
\caption{Distribution of $\beta \Delta U_3$ (Eq.~(\ref{Eq:step_fe_2})). 
The solid line is the Gaussian
model distribution with the sample mean and variance. Positive values
of $\beta
\Delta U_3$ reflect unfavorable crowding of ligands. The required single-point calculations used the PBE functional and the DEF2TZVP basis.}
\label{fig:HOdU3} 
\end{figure}

Evaluation of the energy combination $\Delta U_n$, Eq.~\eqref{Eq:step_fe_2},  
starts with the sampled configuration of the   \ce{(H2O)_n X} cluster. 
Each ligand in turn serves to compose the
energy difference suggested by the exchange 
\begin{multline}
\ce{(H2O)_n X} +
\mathrm{X}  \rightleftharpoons \ce{(H2O)_{n{-1}}X} + \ce{(H2O)X}~. \label{eq:transfer} 
\end{multline}
Geometries of
species on the right of Eq.~\eqref{eq:transfer} conform to the sampled
\ce{(H2O)_n X} structure on the left. To appreciate 
$\Delta U_n$, we use the following accounting.  
Consider first the
contribution $\left\lbrace U\left\lbrack\ce{(H2O)_n X} \right\rbrack -
U\left\lbrack\ce{(H2O)_{n{-}1} X} \right\rbrack\right\rbrace$. This is the
energy change for introducing an additional \ce{H2O} ligand to a
$\ce{(H2O)_{n{-}1}X}$ complex.  The remaining combination $ \left\lbrace
U\left\lbrack\ce{(H2O)X} \right\rbrack -
U\left\lbrack\mathrm{X}\right\rbrack\right\rbrace$ is the energy change for introducing one
\ce{H2O} ligand to a bare X ion. The difference $\Delta U_n$ thus reflects
the crowding of the $n^{\mathrm{th}}$ \ce{H2O} ligand, including any
degradation of binding of the $n^{\mathrm{th}}$ \ce{H2O} ligand to
the X ion (Figure~\ref{fig:HOdU3}). 

Charge is balanced in Eq.~\eqref{eq:2charged}, and the energy combination of Eq.~\eqref{Eq:step_fe_2} is not affected by the electrostatic potential of the phase.\cite{ARPC} 


For $n=1$, Eq.~\eqref{Eq:final} correctly reduces to the trivial case of
$K_{0}{}^{(0)} = 1$. In evaluating $K_{n}{}^{(0)}$ for $n\geq2$,
the value of $K_{1}{}^{(0)}$ can be supplied from experiment
\cite{tissandier1998proton} or alternative theory. This term incorporates the
interaction strength between X and one \ce{H2O} molecule. Carrying out
subsequent steps in this scheme then addresses the issues that make
anion hydration more challenging, \textit{i.e.,}  competing 
interactions of neighboring \ce{H2O} molecules in those clusters.

Figure~\ref{fig:HO_harmonic} shows
that a modest vertical 
shift of those harmonic approximation values substantially 
improves the agreement between harmonic approximation computations and experimental throughout.
Since the approach from Eq.~\eqref{Eq:final} 
takes $K_1^{(0)}$ as input, 
the present more detailed development  should 
benefit similarly. 
This approach does not directly address issues of quantum
mechanical zero-point motion, except to the extent that a pragmatic
inclusion of 
an external $K_1^{(0)}$ incorporates zero-point motion
empirically. 
The empirical grounding 
of this procedure might be particularly 
relevant to the muli-modal possibilities 
of the H-atom that is interior to
\ce{(HO)HOH}$^-$,\cite{marx2010aqueous} 
similar to the Zundel cation 
\ce{(H2O)H(OH2)}$^+$.
This inner-shell evaluation can 
use electronic-structure methods of arbitrary sophistication
since the calculations need be carried-out only for
modest-sized clusters.  These results
re-examine a previous application of QCT to HO$^-$(aq) that 
was strikingly accurate for the dissociation thermodynamics.\cite{asthagiri2003jcp}

\subsubsection{Specific definition of clustering }

In the discussion above, we did not address the
consequences of any specific clustering definition.  That delinquency
amounts to the assumption that
complexes encountered in our cluster simulations,
\textit{i.e.,} at moderate temperatures, 
are well \textit{clustered.} Going further, we
consider how a specific treatment might be built from here. 
A simple clustering criterion is that a clustered water molecule should
donate at least one H atom within an assigned radius
of the ion (Figure~\ref{fig:ClusteringCriterion}). 

We introduce a geometric indicator 
\begin{eqnarray}
\chi_n = \prod\limits_{k\in{ n}} b_{\mathrm{X}}\left( k \right)
\label{eq:clusterindicator}
\end{eqnarray}
of a clustered configuration. $b_{\mathrm{X}}\left( k \right) $ indicates
whether molecule $k$ is clustered (value 1) with the ion X or not (value
0).  $\chi_n$ takes value one if the $n$ molecules are clustered with the
ion, and zero otherwise.  Thus $\chi_n$ is indeed an indicator
function.

We will denote by $\bar{K}_n ^{(0)}$ the equilibrium constant obtained
by our scheme above without definition of a restrictive
inner-shell region. Then
\begin{equation}
    \frac{K_n ^{(0)}}{\bar{K}_n ^{(0)}} = 
    \left\langle \chi_n\right\rangle~,
    \label{eq:Chi}
\end{equation}
where $\left\langle \chi_n\right\rangle$ indicates the
statistical average evaluated as above,
with\textit{\underline{out}} specific definition of
`\textit{clustered.}'

Note that $0\le\chi_n\le1$, and thus $ \ln\left \lbrack K_n^{(0)}/ \bar{K}_n
^{(0)}\right\rbrack \leq 0$. This result makes physical sense because matching the natural clustering should not increase the free energy. 

An interesting physical consideration is that the experimental results
implicitly describe some specific physical clustering. Poorly formed
physical clusters, not-conforming to our mathematically defined
\textit{cluster}, might lose a weakly bound ligand, which would then
populate the outer-shell, perhaps to be pumped-away in the
experiment. These considerations raise the question: ``What
clustering definition is most appropriate for the experiments?''

The estimated average, $\ln \langle \chi_n \rangle$,  sampled from
AIMD dynamics of $\left ( \mathrm{H}_2\mathrm{O} \right )_n
\mathrm{X}$ for $2 \leq n \leq 5$ clusters (Table~\ref{tab:chi_ns}) shows
that, generally, as $n$ increases, $\langle \chi_n \rangle$ decreases as
repulsive interactions between water molecules force ligands to the
outer-shell. This behavior is especially evident for \ce{(H2O)_5I^-} on the one hand, where none
of the sampled configurations had all 5 water molecules within the
defined inner shell.  On the other hand, $\ln \langle \chi_n \rangle \approx 0$
for \ce{(H2O)_3HO^-}
(Table~\ref{tab:chi_ns}), which indicates that the present clustering definition
effectively encompasses the region of physical clustering for that case.
\begin{table*}[ht]
    \centering
\begin{tabular}{|c|c c c c| }\hline
\diagbox{ion (X)}{$n$} & 2 & 3 & 4 & 5\\ \hline
 F$^-$\hfill ($0.27$~nm)  &  0.00 & $-0.21 \pm 0.07$ & $-1.61 \pm 0.28$ & $-1.05\pm 0.19$\\ 
 HO$^-$\hfill ($0.25$~nm) & $-0.02 \pm 0.02$  & 0.00 & $-2.83 \pm 0.56$ & --\\ 
 Cl$^-$\hfill ($0.29$~nm) &  0.00 & $-0.06\pm 0.03$ & $-0.42\pm 0.10$ & $-2.21\pm 0.40$ \\
 Br$^-$\hfill ($0.31$~nm) &  $-0.13\pm 0.05$ & $-0.22\pm 0.07$ & $-0.42\pm 0.10$ & $-1.61\pm 0.28$ \\
 I$^-$ \hfill ($0.34$~nm) &  $-0.13\pm 0.05$ & $-0.36\pm 0.09$ & $-0.20\pm 0.07$ & --\\ \hline
\end{tabular}
\caption{Estimated $ \ln \left\langle \chi_n\right\rangle$, from 
AIMD calculations of \ce{(H2O)_n X} defined 
above. Production
trajectories were sampled at 10/ps, so that the sample sizes were $m=51$. The
indicated uncertainties approximate one standard error according to the
formula $ \sqrt{\left(1 - \left\langle
\chi_n\right\rangle\right)/\left\langle \chi_n\right\rangle m}$, which
assumes independence of the observations. The radius-parameters $R$ (Figure~\ref{fig:ClusteringCriterion})
are noted in
parentheses.  This parameter was chosen to match approximately the first minimum
of the observed $g_{\mathrm{H}|\mathrm{X}}$; see Figure~\ref{fig:HOcompare}. }
    \label{tab:chi_ns}
\end{table*}
    

\subsection{Outer-shell and PCM}

The outer-shell cluster contribution of
Eq.~\eqref{eq:qctX}  to the hydration free energy can be treated\cite{ajay_cl} using the polarizable
continuum model\cite{Tomasi:2005tc} (PCM) in 
the Gaussian suite of electronic structure
programs.\cite{g16}
The  
geometrical structures sampled from the AIMD trajectory for \ce{(H2O)_nX}
were subjected to two single-point electronic calculations, separately; one for the isolated cluster and a second with the external 
(dielectric)
medium described by the PCM tool. The difference, $\bar \varepsilon_j$,  is
employed in computing
\begin{eqnarray}
\mu^{{\mathrm{(ex)}}}_{\mathrm{(H_2O)}_n\mathrm{X}} = -RT
\ln \left\lbrack
\left(\frac{1}{m}\right)\sum\limits_{j=1}^{m}\me^{-{\bar\varepsilon}_j/RT} 
\right\rbrack ,
\label{eq:widom}
\end{eqnarray}
where $m$ is the number of configurations from the simulation stream
that satisfy the clustering definition. Eq.~\eqref{eq:widom} 
corresponds to
the potential distribution theorem (PDT)
approach,\cite{lrp:book,Asthagiri:2010tj} recognizing that
thermal fluctuations implicit in the PCM\cite{Tomasi:2005tc} approach
complete the PDT averaging.

PCM is an approximate description of molecular-scale
aspects of hydration.\cite{pratt1997boundary}   PCM 
does include approximate accounts of 
packing effects and dispersion interactions,
secondarily to 
long-range electrostatic interactions.\cite{Tomasi:2005tc}
Our discussion has emphasized that
QCT is built by identification of inner-shell clusters, separate treatment of
those clusters, then integration of those results into
the broader-scale solution environment. Inevitably,
approximations are required to describe the outer-shell
effects, and here those approximations are bundled in the PCM model.

The approximate
character of the PCM model is signaled by the sensitive
dependence on radii-parameters that locate jumps in dielectric responses 
used in defining the model.   It is reassuring that empirical
values of those parameters are of reasonable magnitude.  Still
the results are sensitive to those radii-parameters, and they are 
not determined by theory\cite{linder1967cavity} or experiment separate 
from the model.  It is striking and 
important that QCT  moderates 
that sensitivity by the subtraction of
the ligand free energies in the formulation
of that outer-shell contribution.\cite{Rempe:2000uw,Asthagiri:2010tj}
That insensitivity is achieved operationally by the boundaries 
of X being somewhat buried by the ligands, and the ligand 
boundaries being unchanged in the subtraction
(Eq.~\eqref{eq:qctX}).  The indicated subtraction
reflects the appearance of the $\rho_{\mathrm{H_2O}}{}^{n}$ factors 
together with $K^{(0)}_{n}$ of 
Eq.~\eqref{eq:qctX}.

For the cases of  F$^-$(aq) and Cl$^-$(aq), we know
already that this PCM-assisted 
application of QCT works satisfactorily,\cite{ajay_cl}
and similarly for HO$^-$(aq).\cite{asthagiri2003jcp,AsthagiriD:ThehsH,gomez2021rough}
For the cases of Br$^-$(aq) and I$^-$(aq), that procedure
is somewhat degraded.\cite{gomez2021rough}  We surmise
that is due to the asymmetric hydration (Figure~\ref{fig:ClusteringCriterion}) of those
ions that leaves the central ion more exposed comparatively.  Therefore, we tried the 
alternative approach,
\begin{eqnarray}
\mu^{{\mathrm{(ex)}}}_{\mathrm{(H_2O)}_n\mathrm{X}} = RT
\ln \left\lbrack
\left(\frac{1}{M}\right)\sum\limits_{j=1}^{M}\me^{{\bar\varepsilon}_j/RT} 
\right\rbrack ,
\label{eq:reversewidom}
\end{eqnarray}
where the $M$ clustered structures are sampled 
from the trajectory of the 
X(aq) AIMD simulations.  This strategy is the 
well-known \underline{\textit{inverse}} 
formula to the standard potential 
distribution theorem,\cite{lrp:book}
but implemented with the PCM tool.
Our physical 
argument is that clustered 
\ce{(H2O)_n X} structures obtained with
this inverse procedure ought to relieve intra-cluster hydrogen bonding and bury the X ion from contact with the water comparatively better (see also Refs.~\citenum{Merchant:2011ga}
\& \citenum{Sabo:2013gs}); thus the severe PCM approximation might 
perform better. Indeed, that was found 
to be the case, 
and the results discussed below (Figure~\ref{fig:hydration_X})
for the cases of Br$^-$(aq) and I$^-$(aq) were obtained 
with this inverse procedure.

It is worth emphasizing the treatment 
of long-range electrostatic interactions  
in this implementation of QCT.  Though 
sampling of the bulk solution structures uses
the preferred and commonplace periodic boundary
conditions in treating  
long-rang interactions, \textit{i.e.} Ewald electrostatics,
the energetics that enter into the free energy computations reported
do not use Ewald electrostatics.

\subsection{Poly-dispersity and net hydration
free energies}

\begin{figure*}[ht]
    \centering
    \includegraphics[width=0.4\textwidth]{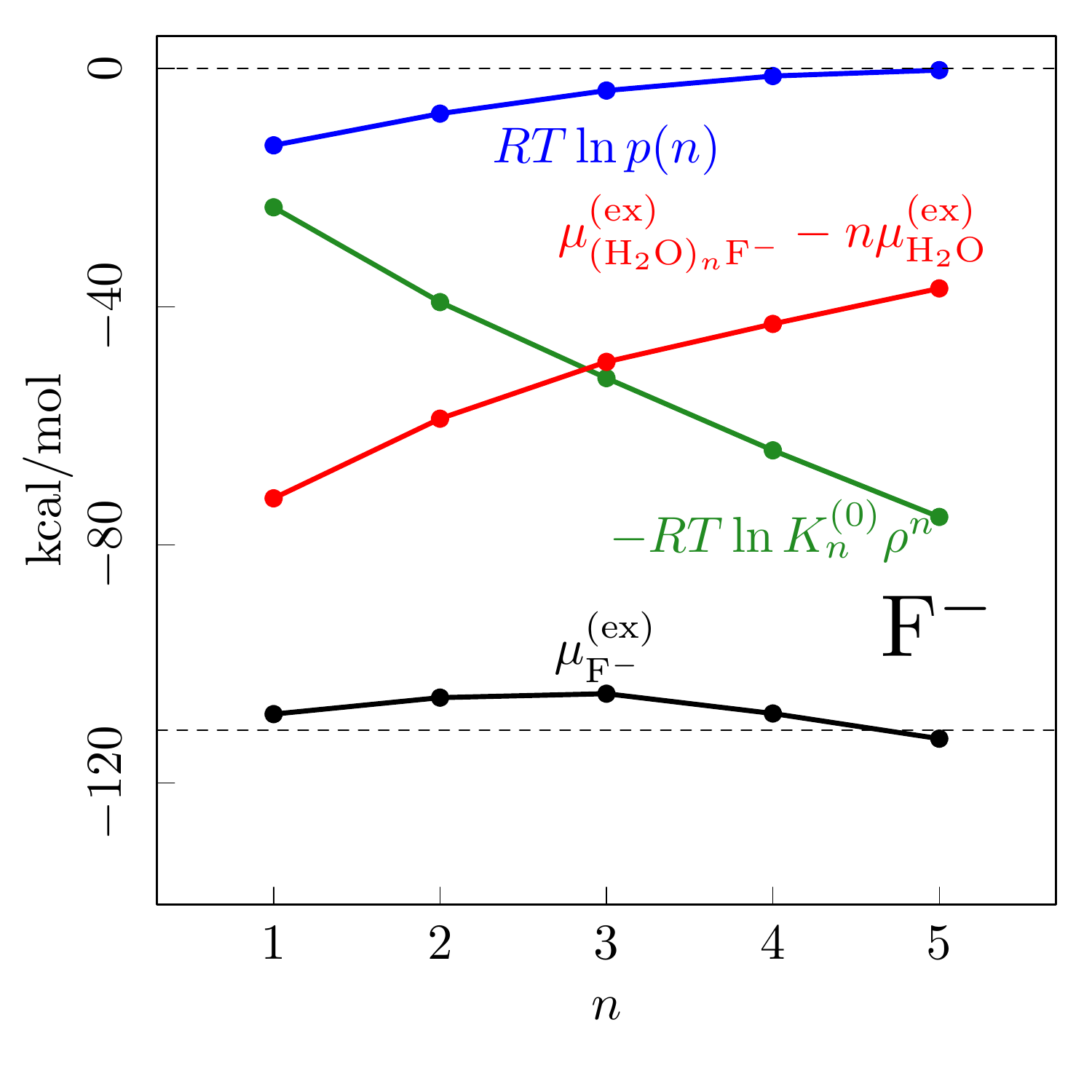}\hfil
    \includegraphics[width=0.4\textwidth]{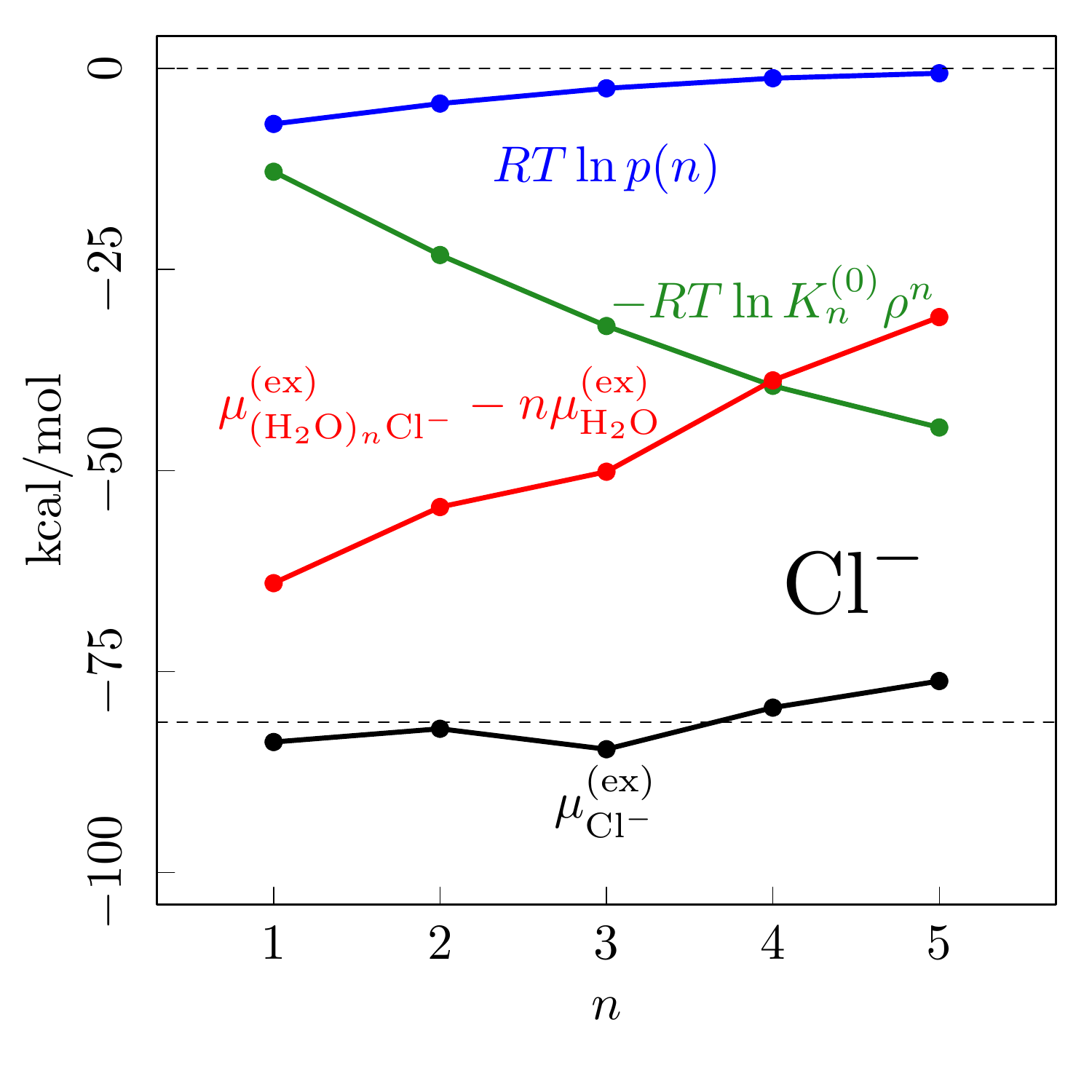}\linebreak
    \includegraphics[width=0.4\textwidth]{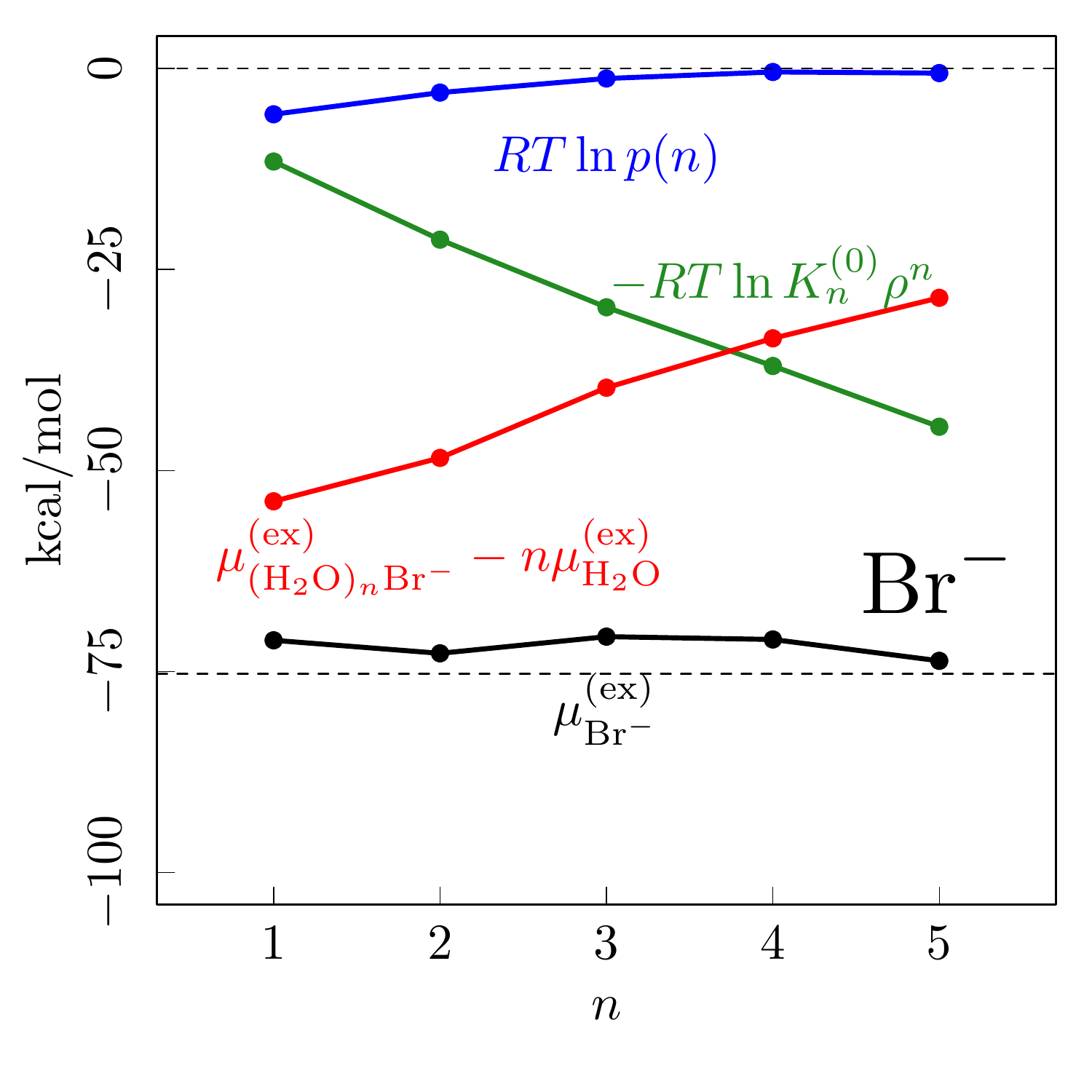}\hfil
    \includegraphics[width=0.4\textwidth]{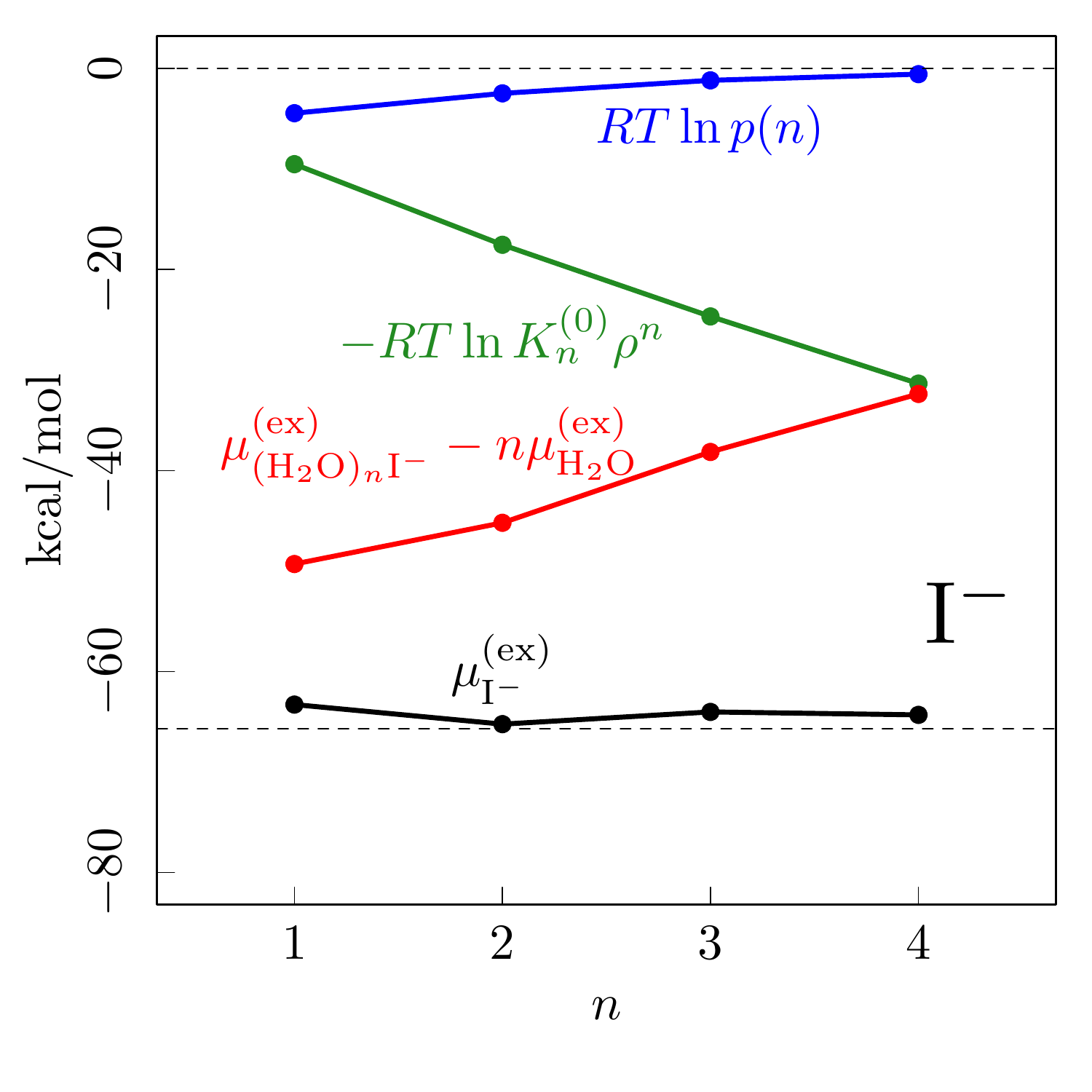}\linebreak
    \caption{Black: The excess hydration free energy of X(aq).
    {\color{Red} Red}: The outer-shell contribution
    evaluated using the PCM \cite{Tomasi:2005tc} with cluster
    configurations sampled from AIMD.  {\color{ForestGreen}
    Green}: Inner-shell free energies.\cite{gomez2021rough} {\color{blue} Blue}: The
    poly-dispersity contribution obtained from the Gaussian model for
    $p(n)$ from AIMD data. In the cases of Br$^-$(aq) and I$^-$(aq),
    these results used the PCM in the
    inverse procedure of Eq.~\eqref{eq:reversewidom}.
    Otherwise, results were obtained with the direct Eq.~\eqref{eq:widom}.}
    \label{fig:hydration_X}
\end{figure*}

\begin{figure*}[ht]
    \centering
    \includegraphics[width=0.4\textwidth]{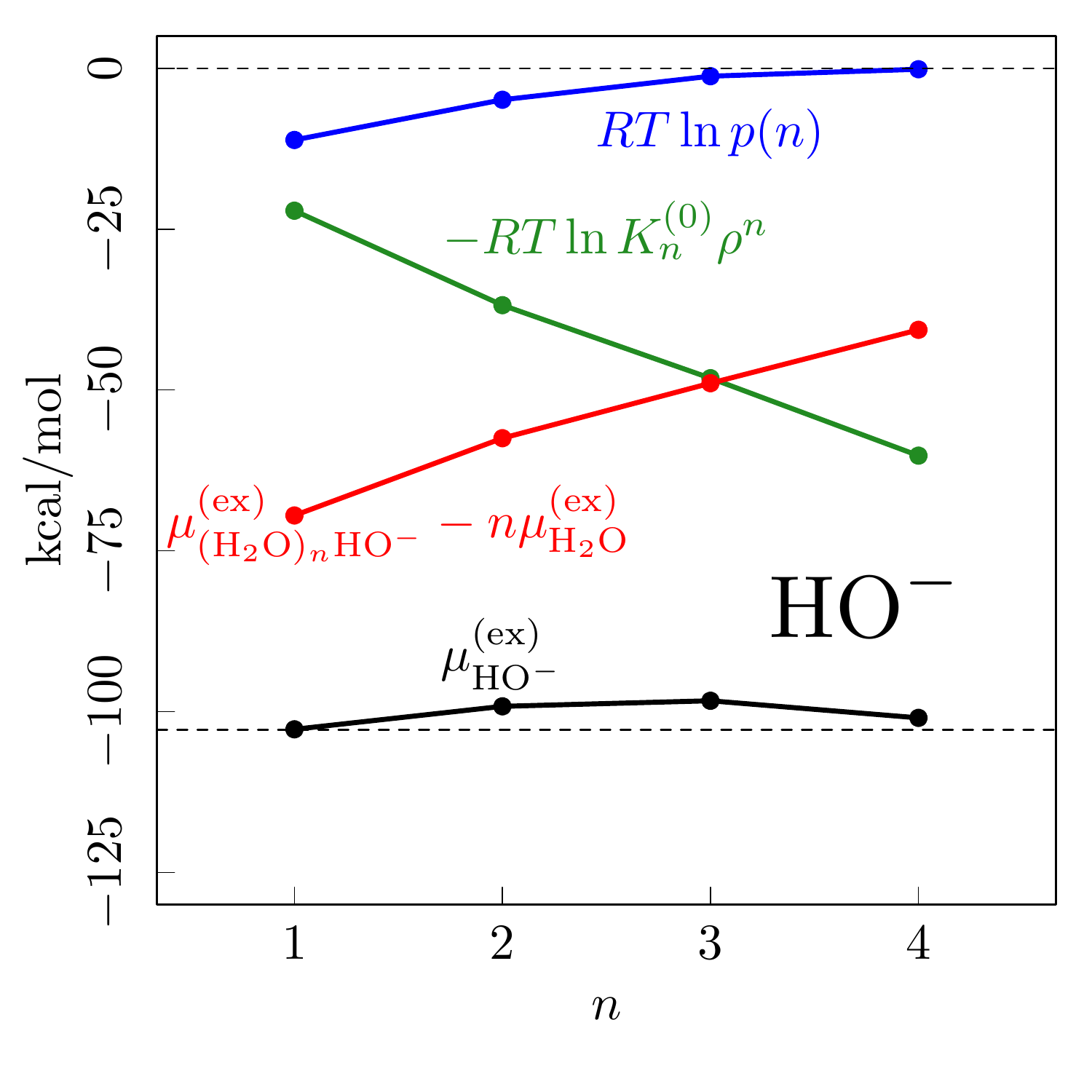}
    \caption{QCT evaluation of hydration free energy
    for HO$^-$(aq), labelled as with Figure~\ref{fig:hydration_X}.}
    \label{fig:hydration_HO}
\end{figure*}

We have noted above that the polydispersity
contribution is the smallest of the three 
contributions to cluster QCT.
It is conceptually simplest, and utilizes 
direct AIMD
simulation in order to estimate
$p_{\mathrm{X}}(n)$.  Thus, at this stage,
we display all three contributions, and their 
combination (Table~\ref{tab:hydration}, 
then Figures~\ref{fig:hydration_X} and
\ref{fig:hydration_HO}),
and then proceed to their physical discussion.
[The reported thermodynamic free energies are obtained from
\begin{multline}
\exp\left(
-\beta\mu_{\mathrm{X}}^{\mathrm{(ex)}}\right)
= 
\sum_n p_{\mathrm{X}}(n) \times \\
\me^{\left\lbrack 
K^{(0)}_n \rho_{\mathrm{H_2O}}{}^n 
-\ln p_{\mathrm{X}}(n) 
-\beta\left(
\mu^{\mathrm{(ex)}}_{\mathrm{(H_2O)_nX}}-n\mu^{\mathrm{(ex)}}_{\mathrm{H_2O}}
\right)
\right\rbrack}~.
\end{multline}
This rearranges Eq.~\protect\eqref{eq:qctX},
then acknowledges the normalization of 
\protect\(p_{\mathrm{X}}(n).\protect\)]

Note again that these free energies
span a chemical scale of energies.   For the least
strongly bound case (I$^-$), the magnitudes of 
the 
net free energies are in excess of 60~kcal/mol,
roughly 100~$RT$ here. 
The net quantities (Eq.~\eqref{eq:qctX}, then 
Figure~\ref{fig:hydration_X} and
\ref{fig:hydration_HO})
are independent of $n$ on that chemical 
energy scale.  The agreement with 
the experimental tabulation (Table~\ref{tab:hydration})
is excellent, and consistent across the anions treated.
The latter point shows that the agreement
is not affected by an assignment of a 
free energy
value for H$^+$(aq) in the experimental tabulation.

\begin{table*}[ht]
\centering
\begin{tabular}{|c c c|}\hline
 X &  $\mu^{ (\mathrm{ex} ) }_{\mathrm{X}}$ (kcal/mol) & Experiment$^\ast$ (kcal/mol)\protect\cite{Marcus:1994ci} \\ \hline 
 F$^-$  &  $-112$ & $-111.1$\\ 
 HO$^-$  &  $-101$ & $-102.8$\\
 Cl$^-$ &  $-82$ & $-81.3$\\ 
 Br$^-$ &  $-73$ & $-75.3$\\ 
 I$^-$  &  $-64$ & $-65.7$\\ \hline
\end{tabular}
\protect\caption{Hydration free energies of anions at standard
conditions. 
The rms difference
of these two columns is approximately 1.6~kcal/mol.
}
\label{tab:hydration}
\end{table*}

\begin{table*}[ht!]
\centering
\begin{tabular}{|c c c|}\hline
 X &  $r_{\mathrm{O}|\mathrm{X}}$ (nm) & Experiment \\ \hline 
 F$^-$  &  $0.268$ & $0.262-0.269$\protect\cite{Ohtaki1993}\\ 
 HO$^-$  &  $0.260$ & $0.265-0.270$\protect\cite{Sipos:2008}\\
 Cl$^-$ &  $0.324$ & $0.310-0.320$\protect\cite{Ohtaki1993}\\ 
 Br$^-$ &  $0.327$ & $0.329-0.340$\protect\cite{Ohtaki1993}\\ 
 I$^-$  &  $0.352$ & $0.350-0.370$\protect\cite{Ohtaki1993,Fulton:I}\\ \hline
\end{tabular}
\caption{Hydration structure of anions: radial position of the 
maximum of the radial distribution function.}
\label{tab:rdf}
\end{table*}

\subsection{Some structural observations}
Though this work has explicitly marshalled simulation calculations
 toward evaluating QCT free energies, some
structural observations are also available.

The cluster definition (Figure~\ref{fig:ClusteringCriterion}
and Table~\ref{tab:chi_ns}) based on H-atom donation, either
singly or doubly, works satisfactorily: small values estimated
for $\ln \left\langle\chi_n\right\rangle$ indicate that the defined clustering volumes
encompass the clustering observed from the AIMD for \ce{(H2O)_n X}
cases, especially for the smaller values of $n$.  For 
example, we see $\ln\left\langle \chi_3 \right\rangle\approx 0$ with \ce{(H2O)_3 HO^-}
(Table~\ref{tab:chi_ns}).  This result
suggests simple H-bond 
donation,  and that  is supported by
the radial 
distribution functions in the \ce{(H2O)_3 HO^-} case 
(Figure~\ref{fig:HOcompare}).  
This observation provides a simple rationale for 
the remarkable success of QCT free energy calculations 
for HO$^-$(aq).\cite{asthagiri2003jcp,AsthagiriD:ThehsH,Asthagiri7229}  
Thus, QCT for hydrated anions works 
better if  the anion inner-shell is characterized by H-bond
donation.  This conclusion is consistent with previous
work,\cite{Chaudhari2017:F,ajay_cl,ajay:F} but focuses 
attention specifically on the XH 
radial distribution functions (Figure~\ref{fig:HOcompare}).

Nevertheless, the structures of the clusters 
are different from 
the bulk aqueous solution case, as depicted by the XH rdfs  (Figure~\ref{fig:HOcompare}).  For the clusters,
setting aside F$^-$ and 
HO$^-$, asymmetric hydration structures 
prevail (Figure~\ref{fig:ClusteringCriterion}).  Asymmetric inner-shell structures 
are moderated in the bulk hydration environment,
but still evident.\cite{Rogers:2010gh} For the bulk hydration cases overall, 
the XH rdfs suggest filling of inner-shells to slightly higher 
average coordination numbers than for the clusters.  Using 
\ce{HO$^-$} again as an example,
the expected coordination number is about 3.8 (Figure~\ref{fig:HOcompare}), 
which may be compared to the 
work of Ref.~\citenum{Asthagiri7229} that estimated 3.7.
The predictions of hydration structure agree well with
experimental estimates based on x-ray and neutron diffraction and x-ray absorption fine structure studies (Table~\ref{tab:rdf}).
Additional information, including traditional XO radial distribution
functions, and further discussion
can be found in the PhD thesis of Diego T. Gomez.\cite{gomez2021rough}

\section{Conclusions}

The final free energies (Figure~\ref{fig:hydration_X}, 
Figure~\ref{fig:hydration_HO}, and Table~\ref{tab:hydration}) are accurate
in comparison with the standard tabulation
Ref.~\citenum{Marcus:1994ci}, and the final composite 
results of 
Figures~\ref{fig:hydration_X} and~\ref{fig:hydration_HO} are independent of 
$n$ on the chemical energy scale of relevance.
 Evaluations of
the inner-shell and polydispersity contributions 
of Eq.~(2) are key to demonstration 
of this theoretical accuracy. Structural information 
obtained by simulations shows that the distinctive asymmetry of
anion clusters is moderated in bulk aqueous solution.

The inner-shell free energy contribution directly tracks available 
experimental information on gas-phase cluster
hydration equilibria,
and the polydispersity contribution is a direct
structural observation from the AIMD trajectory.
Neither of those contributions
is expected to be sensitive to the potential of the phase.

The excellent theory-experiment agreement observed
for those inner-shell
cluster  contributions is a break-through that 
supports the approximate remainder of the theory. 
The $n$-dependent balance 
of  the PCM-approximated, outer-shell  contribution 
with the
remaining numerically exact contributions
suggests
that the PCM approximation performs satisfactorily 
in these applications.  

Implementation
of this PCM-approximate outer-shell contribution
requires
statistical thermodynamic processing for AIMD results
involving single-point electronic structure calculations 
of cluster structures sampled from the AIMD trajectory.
Since the required electronic structure calculations
only treat inner-shell clusters, this electronic
structure effort could employ arbitrarily accurate
numerical theory.  In contrast to earlier 
recommendations, 
further effort 
in that direction is 
not warranted here because of the observed excellent
theory-experiment agreement for the inner-shell cluster
contribution.

Summarizing, the excellent agreement of anion
hydration free energies is due to  quantum computations 
accurately checked with
experiment, \textit{together} with  the physical statistical thermodynamic theory
that enables these computations.  Further, the AIMD simulations reveal differences in anion cluster structures compared with structures found in bulk aqueous solution, with the latter having less asymmetry and higher average coordination numbers. Finally, the ability to predict both accurate solvation free energy and accurate solvation structure of anions supports future work using QCT to understand mechanisms of ion transport and selectivity for the large diversity of anion-selective transport proteins.

\section{Author Information}

\subsection{Authors}

\textbf{Diego T. Gomez}
Department of Chemical \& Biomolecular Engineering, Tulane University, New Orleans LA 70118; https://orcid.org/
0000-0003-0680-6139; Email dgomez1@tulane.edu

\bigskip
\noindent\textbf{Lawrence R. Pratt}
Department of Chemical \& Biomolecular Engineering, Tulane University, New Orleans LA 70118; https://orcid.org/
0000-0003-2351-7451; 
Email lpratt@tulane.edu

\bigskip
\noindent\textbf{Dilipkumar N.  Asthagiri}
Department of Chemical and Biomolecular Engineering, Rice University, Houston TX 77005;  
https://orcid.org/0000-0001-5869-0807; Email dna6@rice.edu

\bigskip
\noindent\textbf{Susan B. Rempe}
Center for Integrated Nanotechnologies, Sandia National Laboratories, Albuquerque, NM 87185; 
https://orcid.org/0000-0003-1623-2108; Email slrempe@sandia.gov

\subsection{Biographies}

\noindent\textbf{Diego T. Gomez} was born in Albuquerque, New Mexico. He holds Chemical Engineering degrees from New Mexico State University (BS, 2017) and Tulane University (PhD, 2021). His contributions include quasi-chemical theory for aqueous halide solutions and theoretical studies of capillary bridges.

\bigskip
\noindent\textbf{Lawrence R. Pratt} was born in Flint, Michigan. He is Professor Emeritus of Chemical Engineering at  Tulane University. He holds Chemistry degrees from Michigan State University (BS, 1972) and the University of Illinois (MS, 1974 and PhD, 1977). Prior to Tulane, he spent 23 years as a Technical Staff Member at Los Alamos National Laboratory. He has contributed to the theory of the hydrophobic effect, the development of transition path sampling, contributions to orbital free density functional theory, and the theory of liquids and solutions.  

\bigskip\noindent\textbf{Dilipkumar N. Asthagiri} was born in Trichy, India. He is an Associate Research Professor of Chemical Engineering at  Rice University.  He holds degrees in Chemical Engineering from 
the University of Delaware (PhD, 1999), University of Michigan (MS, 1994), and the Indian Institute of Technology-Madras (BTech, 1992). His contributions include those in the molecular statistical thermodynamics of hydration and its role in bio-molecular structure and assembly, and NMR relaxation of confined fluids and MRI contrast agents.

\bigskip\noindent\textbf{Susan B. Rempe} was born in Spokane, WA and grew up in Kalispell, MT. She is a Senior Scientist
at Sandia National Labs, Affiliate Scientist at the Center for Integrated Nanotechnologies, and Research Professor at the University of New Mexico. She holds degrees in
Chemistry from the University of Washington (PhD, 1998; 
Masters, 1993) and University of Montana (BA, 1989), and
a degree in Pre-Medical Sciences from Columbia University (BA, 1987). She has contributed to the theory of liquids and condensed matter,  selective ion binding and transport, hydrophobic effects, and the design of separation membranes.

\begin{acknowledgement}
This work was performed, in part, at the
Center for Integrated Nanotechnologies, an Office of Science User
Facility operated for the U.S. Department of Energy (DOE) Office of
Science. The work was supported by Sandia National Laboratories (SNL)
LDRD program. SNL is a multi-mission laboratory managed and operated by
National Technology and Engineering Solutions of Sandia, LLC., a wholly
owned subsidiary of Honeywell International, Inc., for the U.S. DOE's
National Nuclear Security Administration under contract DE-NA-0003525.
The views expressed in the article do not necessarily represent the
views of the U.S. DOE or the United States Government. 

\end{acknowledgement}

\newpage



\providecommand{\latin}[1]{#1}
\makeatletter
\providecommand{\doi}
  {\begingroup\let\do\@makeother\dospecials
  \catcode`\{=1 \catcode`\}=2 \doi@aux}
\providecommand{\doi@aux}[1]{\endgroup\texttt{#1}}
\makeatother
\providecommand*\mcitethebibliography{\thebibliography}
\csname @ifundefined\endcsname{endmcitethebibliography}
  {\let\endmcitethebibliography\endthebibliography}{}
\begin{mcitethebibliography}{95}
\providecommand*\natexlab[1]{#1}
\providecommand*\mciteSetBstSublistMode[1]{}
\providecommand*\mciteSetBstMaxWidthForm[2]{}
\providecommand*\mciteBstWouldAddEndPuncttrue
  {\def\EndOfBibitem{\unskip.}}
\providecommand*\mciteBstWouldAddEndPunctfalse
  {\let\EndOfBibitem\relax}
\providecommand*\mciteSetBstMidEndSepPunct[3]{}
\providecommand*\mciteSetBstSublistLabelBeginEnd[3]{}
\providecommand*\EndOfBibitem{}
\mciteSetBstSublistMode{f}
\mciteSetBstMaxWidthForm{subitem}{(\alph{mcitesubitemcount})}
\mciteSetBstSublistLabelBeginEnd
  {\mcitemaxwidthsubitemform\space}
  {\relax}
  {\relax}

\bibitem[Asthagiri \latin{et~al.}(2010)Asthagiri, Dixit, Merchant, Paulaitis,
  Pratt, Rempe, and Varma]{Asthagiri:2010tj}
Asthagiri,~D.; Dixit,~P.; Merchant,~S.; Paulaitis,~M.; Pratt,~L.; Rempe,~S.;
  Varma,~S. {Ion selectivity from local configurations of ligands in solutions
  and ion channels}. \emph{Chem. Phys. Lett.} \textbf{2010}, \emph{485},
  1--7\relax
\mciteBstWouldAddEndPuncttrue
\mciteSetBstMidEndSepPunct{\mcitedefaultmidpunct}
{\mcitedefaultendpunct}{\mcitedefaultseppunct}\relax
\EndOfBibitem
\bibitem[Gomez \latin{et~al.}(2021)Gomez, Pratt, Rogers, and
  Rempe]{gomez2021free}
Gomez,~D.~T.; Pratt,~L.~R.; Rogers,~D.~M.; Rempe,~S.~B. Free Energies of
  Hydrated Halide Anions: High Through-Put Computations on Clusters to Treat
  Rough Energy-Landscapes. \emph{Molecules} \textbf{2021}, \emph{26},
  3087\relax
\mciteBstWouldAddEndPuncttrue
\mciteSetBstMidEndSepPunct{\mcitedefaultmidpunct}
{\mcitedefaultendpunct}{\mcitedefaultseppunct}\relax
\EndOfBibitem
\bibitem[Muralidharan \latin{et~al.}(2019)Muralidharan, Pratt, Chaudhari, and
  Rempe]{ajay_cl}
Muralidharan,~A.; Pratt,~L.~R.; Chaudhari,~M.~I.; Rempe,~S.~B. {Quasi-chemical
  theory for anion hydration and specific ion effects: Cl$^-$(aq) vs.
  F$^-$(aq)}. \emph{Chem. Phys. Lett.: X} \textbf{2019}, \emph{4}, 100037\relax
\mciteBstWouldAddEndPuncttrue
\mciteSetBstMidEndSepPunct{\mcitedefaultmidpunct}
{\mcitedefaultendpunct}{\mcitedefaultseppunct}\relax
\EndOfBibitem
\bibitem[Chaudhari \latin{et~al.}(2020)Chaudhari, Vanegas, Pratt, Muralidharan,
  and Rempe]{ARPC}
Chaudhari,~M.~I.; Vanegas,~J.~M.; Pratt,~L.; Muralidharan,~A.; Rempe,~S.~B.
  Hydration mimicry by membrane ion channels. \emph{Ann. Rev. Phys. Chem.}
  \textbf{2020}, \emph{71}, 461--484\relax
\mciteBstWouldAddEndPuncttrue
\mciteSetBstMidEndSepPunct{\mcitedefaultmidpunct}
{\mcitedefaultendpunct}{\mcitedefaultseppunct}\relax
\EndOfBibitem
\bibitem[Castleman and Keesee(1986)Castleman, and Keesee]{Castleman:1986fu}
Castleman,~A.~W.; Keesee,~R.~G. {Ionic clusters}. \emph{Chem. Rev.}
  \textbf{1986}, \emph{86}, 589 -- 618\relax
\mciteBstWouldAddEndPuncttrue
\mciteSetBstMidEndSepPunct{\mcitedefaultmidpunct}
{\mcitedefaultendpunct}{\mcitedefaultseppunct}\relax
\EndOfBibitem
\bibitem[Keesee and Castleman~Jr(1986)Keesee, and
  Castleman~Jr]{keesee1986thermochemical}
Keesee,~R.; Castleman~Jr,~A. Thermochemical data on gas-phase ion-molecule
  association and clustering reactions. \emph{J. Phys. {\&} Chem. Ref. Data}
  \textbf{1986}, \emph{15}, 1011--1071\relax
\mciteBstWouldAddEndPuncttrue
\mciteSetBstMidEndSepPunct{\mcitedefaultmidpunct}
{\mcitedefaultendpunct}{\mcitedefaultseppunct}\relax
\EndOfBibitem
\bibitem[Pratt and Rempe(1999)Pratt, and Rempe]{pratt1999quasi}
Pratt,~L.~R.; Rempe,~S.~B. Quasi-Chemical Theory and Implicit Solvent Models
  for Simulations. Simulation and Theory of Electrostatic Interactions in
  Solution. Computational Chemistry, Biophysics, and Aqueous Solutions, Vol.
  492 of AIP Conference Proceedings, American Institute of Physics. 1999; pp
  172--201\relax
\mciteBstWouldAddEndPuncttrue
\mciteSetBstMidEndSepPunct{\mcitedefaultmidpunct}
{\mcitedefaultendpunct}{\mcitedefaultseppunct}\relax
\EndOfBibitem
\bibitem[Asthagiri \latin{et~al.}(2021)Asthagiri, Paulaitis, and
  Pratt]{APP:jpcb21perspective}
Asthagiri,~D.~N.; Paulaitis,~M.~E.; Pratt,~L.~R. {Thermodynamics of Hydration
  from the Perspective of the Molecular Quasichemical Theory of Solutions}.
  \emph{J. Phys. Chem. B} \textbf{2021}, \emph{125}, 8294--8304\relax
\mciteBstWouldAddEndPuncttrue
\mciteSetBstMidEndSepPunct{\mcitedefaultmidpunct}
{\mcitedefaultendpunct}{\mcitedefaultseppunct}\relax
\EndOfBibitem
\bibitem[Kunz(2010)]{kunz2010specific}
Kunz,~W. {Specific ion effects in colloidal and biological systems}.
  \emph{Curr. Opin. Colloid Interface Sci} \textbf{2010}, \emph{15},
  34--39\relax
\mciteBstWouldAddEndPuncttrue
\mciteSetBstMidEndSepPunct{\mcitedefaultmidpunct}
{\mcitedefaultendpunct}{\mcitedefaultseppunct}\relax
\EndOfBibitem
\bibitem[Zhang and Cremer(2010)Zhang, and Cremer]{Zhang:2010gr}
Zhang,~Y.; Cremer,~P.~S. {Chemistry of Hofmeister Anions and Osmolytes}.
  \emph{Annu Rev Phys Chem} \textbf{2010}, \emph{61}, 63--83\relax
\mciteBstWouldAddEndPuncttrue
\mciteSetBstMidEndSepPunct{\mcitedefaultmidpunct}
{\mcitedefaultendpunct}{\mcitedefaultseppunct}\relax
\EndOfBibitem
\bibitem[Pollard and Beck(2016)Pollard, and Beck]{Pollard:2016ei}
Pollard,~T.~P.; Beck,~T.~L. {Toward a quantitative theory of Hofmeister
  phenomena: From quantum effects to thermodynamics}. \emph{Curr. Opin. Coll.
  {\&} Interf. Sci.} \textbf{2016}, \emph{23}, 110--118\relax
\mciteBstWouldAddEndPuncttrue
\mciteSetBstMidEndSepPunct{\mcitedefaultmidpunct}
{\mcitedefaultendpunct}{\mcitedefaultseppunct}\relax
\EndOfBibitem
\bibitem[Rogers \latin{et~al.}(2013)Rogers, Jiao, Pratt, and Rempe]{Rogers}
Rogers,~D.~M.; Jiao,~D.; Pratt,~L.; Rempe,~S.~B. Structural Models and
  Molecular Thermodynamics of Hydration of Ions and Small Molecules. \emph{Ann.
  Rep. Comp. Chem.} \textbf{2013}, \emph{8}, 71--128\relax
\mciteBstWouldAddEndPuncttrue
\mciteSetBstMidEndSepPunct{\mcitedefaultmidpunct}
{\mcitedefaultendpunct}{\mcitedefaultseppunct}\relax
\EndOfBibitem
\bibitem[Chaudhari \latin{et~al.}(2017)Chaudhari, Rempe, and
  Pratt]{Chaudhari2017:F}
Chaudhari,~M.~I.; Rempe,~S.~B.; Pratt,~L.~R. {Quasi-chemical theory of F {$^-$}
  (aq): The ``no split occupancies rule'' revisited}. \emph{J. Chem. Phys.}
  \textbf{2017}, \emph{147}, 161728 -- 5\relax
\mciteBstWouldAddEndPuncttrue
\mciteSetBstMidEndSepPunct{\mcitedefaultmidpunct}
{\mcitedefaultendpunct}{\mcitedefaultseppunct}\relax
\EndOfBibitem
\bibitem[Muralidharan \latin{et~al.}(2018)Muralidharan, Pratt, Chaudhari, and
  Rempe]{ajay:F}
Muralidharan,~A.; Pratt,~L.~R.; Chaudhari,~M.~I.; Rempe,~S.~B. Quasi-Chemical
  Theory With Cluster Sampling From \emph{Ab Initio} Molecular Dynamics:
  Fluoride (F$^-$(aq)) Anion Hydration. \emph{J. Phys. Chem. A} \textbf{2018},
  \emph{122}, 9806--9812\relax
\mciteBstWouldAddEndPuncttrue
\mciteSetBstMidEndSepPunct{\mcitedefaultmidpunct}
{\mcitedefaultendpunct}{\mcitedefaultseppunct}\relax
\EndOfBibitem
\bibitem[Asthagiri \latin{et~al.}(2004)Asthagiri, Pratt, Kress, and
  Gomez]{Asthagiri7229}
Asthagiri,~D.; Pratt,~L.~R.; Kress,~J.~D.; Gomez,~M.~A. Hydration and mobility
  of HO{$^-$}(aq). \emph{Proc. Natl. Acad. Sci. USA} \textbf{2004}, \emph{101},
  7229--7233\relax
\mciteBstWouldAddEndPuncttrue
\mciteSetBstMidEndSepPunct{\mcitedefaultmidpunct}
{\mcitedefaultendpunct}{\mcitedefaultseppunct}\relax
\EndOfBibitem
\bibitem[Marx \latin{et~al.}(2010)Marx, Chandra, and
  Tuckerman]{marx2010aqueous}
Marx,~D.; Chandra,~A.; Tuckerman,~M.~E. {Aqueous Basic Solutions: Hydroxide
  Solvation, Structural Diffusion, and Comparison to the Hydrated Proton}.
  \emph{Chem. Rev.} \textbf{2010}, \emph{110}, 2174--2216\relax
\mciteBstWouldAddEndPuncttrue
\mciteSetBstMidEndSepPunct{\mcitedefaultmidpunct}
{\mcitedefaultendpunct}{\mcitedefaultseppunct}\relax
\EndOfBibitem
\bibitem[Agmon \latin{et~al.}(2016)Agmon, Bakker, Campen, Henchman, Pohl, Roke,
  Thämer, and Hassanali]{Agmon2016}
Agmon,~N.; Bakker,~H.~J.; Campen,~R.~K.; Henchman,~R.~H.; Pohl,~P.; Roke,~S.;
  Thämer,~M.; Hassanali,~A. {Protons and Hydroxide Ions in Aqueous Systems.}
  \emph{Chem. Rev.} \textbf{2016}, \emph{116}, 7642--72\relax
\mciteBstWouldAddEndPuncttrue
\mciteSetBstMidEndSepPunct{\mcitedefaultmidpunct}
{\mcitedefaultendpunct}{\mcitedefaultseppunct}\relax
\EndOfBibitem
\bibitem[Glusker \latin{et~al.}(1999)Glusker, Katz, and Bock]{Glusker:1999}
Glusker,~J.~P.; Katz,~A.~K.; Bock,~C. METAL IONS IN BIOLOGICAL SYSTEMS.
  1999\relax
\mciteBstWouldAddEndPuncttrue
\mciteSetBstMidEndSepPunct{\mcitedefaultmidpunct}
{\mcitedefaultendpunct}{\mcitedefaultseppunct}\relax
\EndOfBibitem
\bibitem[Thomson and Gray(1998)Thomson, and Gray]{Gray}
Thomson,~A.~J.; Gray,~H.~B. Bio-inorganic chemistry. \emph{Curr. Op. Chem.
  Bio.} \textbf{1998}, \emph{2}, 155--158\relax
\mciteBstWouldAddEndPuncttrue
\mciteSetBstMidEndSepPunct{\mcitedefaultmidpunct}
{\mcitedefaultendpunct}{\mcitedefaultseppunct}\relax
\EndOfBibitem
\bibitem[Valdez \latin{et~al.}(2014)Valdez, Smith, Nechay, and
  Alexandrova]{mysteries}
Valdez,~C.~E.; Smith,~Q.~A.; Nechay,~M.~R.; Alexandrova,~A.~N. Mysteries of
  Metals in Metalloenzymes. \emph{Acc. Chem. Res.} \textbf{2014}, \emph{47},
  3110--3117, PMID: 25207938\relax
\mciteBstWouldAddEndPuncttrue
\mciteSetBstMidEndSepPunct{\mcitedefaultmidpunct}
{\mcitedefaultendpunct}{\mcitedefaultseppunct}\relax
\EndOfBibitem
\bibitem[Hille(2001)]{hille}
Hille,~B. \emph{{Ionic Channels of Excitable Membranes}}, 3rd ed.; Sinauer
  Associates: Sunderland, MA, 2001\relax
\mciteBstWouldAddEndPuncttrue
\mciteSetBstMidEndSepPunct{\mcitedefaultmidpunct}
{\mcitedefaultendpunct}{\mcitedefaultseppunct}\relax
\EndOfBibitem
\bibitem[Wulff and Zhorov(2008)Wulff, and Zhorov]{Wulff}
Wulff,~H.; Zhorov,~B.~S. K$^+$ Channel Modulators for the Treatment of
  Neurological Disorders and Autoimmune Diseases. \emph{Chem. Rev.}
  \textbf{2008}, 1744--73\relax
\mciteBstWouldAddEndPuncttrue
\mciteSetBstMidEndSepPunct{\mcitedefaultmidpunct}
{\mcitedefaultendpunct}{\mcitedefaultseppunct}\relax
\EndOfBibitem
\bibitem[Pardo and Stuhmer(2014)Pardo, and Stuhmer]{Pardo}
Pardo,~L.~A.; Stuhmer,~W. The Roles of K$^+$ Channels in Cancer{}. \emph{Nat.
  Rev. Cancer.} \textbf{2014}, 39--48\relax
\mciteBstWouldAddEndPuncttrue
\mciteSetBstMidEndSepPunct{\mcitedefaultmidpunct}
{\mcitedefaultendpunct}{\mcitedefaultseppunct}\relax
\EndOfBibitem
\bibitem[Morales-L{\'a}zaro \latin{et~al.}(2015)Morales-L{\'a}zaro,
  Hern{\'a}ndez-Garc{\'\i}a, Serrano-Flores, and Rosenbaum]{Rosenbaum}
Morales-L{\'a}zaro,~S.~L.; Hern{\'a}ndez-Garc{\'\i}a,~E.; Serrano-Flores,~B.;
  Rosenbaum,~T. {Organic Toxins as Tools to Understand Ion Channel Mechanisms
  and Structure.} \emph{Curr. Top. Med. Chem.} \textbf{2015}, \emph{15},
  581--603\relax
\mciteBstWouldAddEndPuncttrue
\mciteSetBstMidEndSepPunct{\mcitedefaultmidpunct}
{\mcitedefaultendpunct}{\mcitedefaultseppunct}\relax
\EndOfBibitem
\bibitem[Alvarez-Leefmans and Delpire(2009)Alvarez-Leefmans, and
  Delpire]{alvarez2009physiology}
Alvarez-Leefmans,~F.~J.; Delpire,~E. \emph{Physiology and Pathology of Chloride
  Transporters and Channels in the Nervous System: From Molecules to Diseases};
  Academic Press, 2009\relax
\mciteBstWouldAddEndPuncttrue
\mciteSetBstMidEndSepPunct{\mcitedefaultmidpunct}
{\mcitedefaultendpunct}{\mcitedefaultseppunct}\relax
\EndOfBibitem
\bibitem[Kim \latin{et~al.}(2016)Kim, Kwon, Jun, Cha, Kim, Lee, Kim, and
  Choo]{Kim:2016}
Kim,~K.; Kwon,~S.-K.; Jun,~S.-H.; Cha,~J.~S.; Kim,~H.; Lee,~W.; Kim,~J.;
  Choo,~H.-S. {Crystal structure and functional characterization of a
  light-driven chloride pump having an NTQ motif}. \emph{Nat. Commun.}
  \textbf{2016}, \emph{7}, 12677\relax
\mciteBstWouldAddEndPuncttrue
\mciteSetBstMidEndSepPunct{\mcitedefaultmidpunct}
{\mcitedefaultendpunct}{\mcitedefaultseppunct}\relax
\EndOfBibitem
\bibitem[VanGordon \latin{et~al.}(2021)VanGordon, Prignano, Dempski, Rick, and
  Rempe]{vanGordon:2021}
VanGordon,~M.~R.; Prignano,~L.~A.; Dempski,~R.~E.; Rick,~S.~W.; Rempe,~S.~B.
  {Channelrhodopsin C1C2: Photocycle kinetics and interactions near the central
  gate}. \emph{Biophys. J.} \textbf{2021}, \emph{120}, 1835 -- 1845\relax
\mciteBstWouldAddEndPuncttrue
\mciteSetBstMidEndSepPunct{\mcitedefaultmidpunct}
{\mcitedefaultendpunct}{\mcitedefaultseppunct}\relax
\EndOfBibitem
\bibitem[Stockbridge \latin{et~al.}(2015)Stockbridge, Kolmakova-Partensky,
  Shane, Koide, Koide, Miller, and Newstead]{stockbridge2015crystal}
Stockbridge,~R.~B.; Kolmakova-Partensky,~L.; Shane,~T.; Koide,~A.; Koide,~S.;
  Miller,~C.; Newstead,~S. Crystal Structures of a Double-Barrelled Fluoride
  Ion Channel. \emph{Nature} \textbf{2015}, \emph{525}, 548\relax
\mciteBstWouldAddEndPuncttrue
\mciteSetBstMidEndSepPunct{\mcitedefaultmidpunct}
{\mcitedefaultendpunct}{\mcitedefaultseppunct}\relax
\EndOfBibitem
\bibitem[Varma and Rempe(2007)Varma, and Rempe]{varma2007tuning}
Varma,~S.; Rempe,~S.~B. Tuning ion coordination architectures to enable
  selective partitioning. \emph{Biophys. J.} \textbf{2007}, \emph{93},
  1093--1099\relax
\mciteBstWouldAddEndPuncttrue
\mciteSetBstMidEndSepPunct{\mcitedefaultmidpunct}
{\mcitedefaultendpunct}{\mcitedefaultseppunct}\relax
\EndOfBibitem
\bibitem[Varma \latin{et~al.}(2008)Varma, Sabo, and
  Rempe]{varma2008valinomycin}
Varma,~S.; Sabo,~D.; Rempe,~S.~B. {K$^+$/Na$^+$} Selectivity in {K} Channels
  and Valinomycin: Over-coordination Versus Cavity-size constraints. \emph{J.
  Mol. Biol.} \textbf{2008}, \emph{376}, 13--22\relax
\mciteBstWouldAddEndPuncttrue
\mciteSetBstMidEndSepPunct{\mcitedefaultmidpunct}
{\mcitedefaultendpunct}{\mcitedefaultseppunct}\relax
\EndOfBibitem
\bibitem[Fowler \latin{et~al.}(2008)Fowler, Tai, and Sansom]{Fowler2008}
Fowler,~P.; Tai,~K.; Sansom,~M. The selectivity of {K$^+$} ion channels:
  Testing the hypotheses. \emph{Biophys. J.} \textbf{2008}, \emph{95},
  5062--5072\relax
\mciteBstWouldAddEndPuncttrue
\mciteSetBstMidEndSepPunct{\mcitedefaultmidpunct}
{\mcitedefaultendpunct}{\mcitedefaultseppunct}\relax
\EndOfBibitem
\bibitem[Varma and Rempe(2008)Varma, and Rempe]{Varma:2008jacs}
Varma,~S.; Rempe,~S.~B. Structural Transitions in Ion Coordination Driven by
  Changes in Competition for Ligand Binding. \emph{J. Am. Chem. Soc.}
  \textbf{2008}, \emph{130}, 15405--15419, PMID: 18954053\relax
\mciteBstWouldAddEndPuncttrue
\mciteSetBstMidEndSepPunct{\mcitedefaultmidpunct}
{\mcitedefaultendpunct}{\mcitedefaultseppunct}\relax
\EndOfBibitem
\bibitem[Varma and Rempe(2010)Varma, and Rempe]{Varma:2010}
Varma,~S.; Rempe,~S.~B. {Multibody Effects in Ion Binding and Selectivity}.
  \emph{Biophys. J.} \textbf{2010}, \emph{99}, 3394--3401\relax
\mciteBstWouldAddEndPuncttrue
\mciteSetBstMidEndSepPunct{\mcitedefaultmidpunct}
{\mcitedefaultendpunct}{\mcitedefaultseppunct}\relax
\EndOfBibitem
\bibitem[Furini and Domene(2011)Furini, and Domene]{furini2011}
Furini,~S.; Domene,~C. Selectivity and Permeation of Alkali Metal Ions in
  {K$^+$}-channels. \emph{J. Mol. Biol.} \textbf{2011}, \emph{409},
  867--878\relax
\mciteBstWouldAddEndPuncttrue
\mciteSetBstMidEndSepPunct{\mcitedefaultmidpunct}
{\mcitedefaultendpunct}{\mcitedefaultseppunct}\relax
\EndOfBibitem
\bibitem[Rossi \latin{et~al.}(2013)Rossi, Tkatchenko, Rempe, and Varma]{rossi}
Rossi,~M.; Tkatchenko,~A.; Rempe,~S.~B.; Varma,~S. {Role of Methyl-Induced
  Polarization in Ion Binding}. \emph{Proc. Natl. Acad. Sci. USA}
  \textbf{2013}, \emph{110}, 12978--12983\relax
\mciteBstWouldAddEndPuncttrue
\mciteSetBstMidEndSepPunct{\mcitedefaultmidpunct}
{\mcitedefaultendpunct}{\mcitedefaultseppunct}\relax
\EndOfBibitem
\bibitem[Medovoy \latin{et~al.}(2016)Medovoy, Perozo, and Roux]{medovoy:2016kr}
Medovoy,~D.; Perozo,~E.; Roux,~B. {Multi-ion free energy landscapes underscore
  the microscopic mechanism of ion selectivity in the KcsA channel}.
  \emph{BBA-Biomembranes} \textbf{2016}, \emph{1858}, 1722--1732\relax
\mciteBstWouldAddEndPuncttrue
\mciteSetBstMidEndSepPunct{\mcitedefaultmidpunct}
{\mcitedefaultendpunct}{\mcitedefaultseppunct}\relax
\EndOfBibitem
\bibitem[Kopec \latin{et~al.}(2018)Kopec, Kopfer, Vickery, Bondarenko, Jansen,
  de~Groot, and Zachariae]{deGroot2018}
Kopec,~W.; Kopfer,~D.; Vickery,~O.; Bondarenko,~A.~S.; Jansen,~T. L.~C.;
  de~Groot,~B.~L.; Zachariae,~U. Direct knock-on of desolvated ions governs
  strict ion selectivity in {K$^+$} channels. \emph{Nat. Chem.} \textbf{2018},
  \emph{10}, 813--820\relax
\mciteBstWouldAddEndPuncttrue
\mciteSetBstMidEndSepPunct{\mcitedefaultmidpunct}
{\mcitedefaultendpunct}{\mcitedefaultseppunct}\relax
\EndOfBibitem
\bibitem[Jing \latin{et~al.}(2021)Jing, Rackers, Pratt, Liu, Rempe, and
  Ren]{jing2021thermodynamics}
Jing,~Z.; Rackers,~J.~A.; Pratt,~L.; Liu,~C.; Rempe,~S.~B.; Ren,~P.
  Thermodynamics of ion binding and occupancy in potassium channels.
  \emph{Chem. Sci.} \textbf{2021}, \relax
\mciteBstWouldAddEndPunctfalse
\mciteSetBstMidEndSepPunct{\mcitedefaultmidpunct}
{}{\mcitedefaultseppunct}\relax
\EndOfBibitem
\bibitem[Ko and Jo(2010)Ko, and Jo]{ko2010chloride}
Ko,~Y.~J.; Jo,~W.~H. Chloride ion conduction without water coordination in the
  pore of {ClC} protein. \emph{J. Comp. Chem.} \textbf{2010}, \emph{31},
  603--611\relax
\mciteBstWouldAddEndPuncttrue
\mciteSetBstMidEndSepPunct{\mcitedefaultmidpunct}
{\mcitedefaultendpunct}{\mcitedefaultseppunct}\relax
\EndOfBibitem
\bibitem[Kuang \latin{et~al.}(2008)Kuang, Liu, and Beck]{kuang2008transpath}
Kuang,~Z.; Liu,~A.; Beck,~T.~L. TransPath: A computational method for locating
  ion transit pathways through membrane proteins. \emph{Proteins}
  \textbf{2008}, \emph{71}, 1349--1359\relax
\mciteBstWouldAddEndPuncttrue
\mciteSetBstMidEndSepPunct{\mcitedefaultmidpunct}
{\mcitedefaultendpunct}{\mcitedefaultseppunct}\relax
\EndOfBibitem
\bibitem[Yin \latin{et~al.}(2004)Yin, Kuang, Mahankali, and Beck]{yin2004ion}
Yin,~J.; Kuang,~Z.; Mahankali,~U.; Beck,~T.~L. Ion transit pathways and gating
  in {ClC} chloride channels. \emph{Proteins: Structure, Function, and
  Bioinformatics} \textbf{2004}, \emph{57}, 414--421\relax
\mciteBstWouldAddEndPuncttrue
\mciteSetBstMidEndSepPunct{\mcitedefaultmidpunct}
{\mcitedefaultendpunct}{\mcitedefaultseppunct}\relax
\EndOfBibitem
\bibitem[Chen and Beck(2016)Chen, and Beck]{chen2016free}
Chen,~Z.; Beck,~T.~L. Free energies of ion binding in the bacterial {ClC-Ec1}
  chloride transporter with implications for the transport mechanism and
  selectivity. \emph{J. Phys. Chem. B B} \textbf{2016}, \emph{120},
  3129--3139\relax
\mciteBstWouldAddEndPuncttrue
\mciteSetBstMidEndSepPunct{\mcitedefaultmidpunct}
{\mcitedefaultendpunct}{\mcitedefaultseppunct}\relax
\EndOfBibitem
\bibitem[Yue \latin{et~al.}(2022)Yue, Wang, and Voth]{Voth:2022}
Yue,~Z.; Wang,~Z.; Voth,~G. {Ion permeation, selectivity, and electronic
  polarization in fluoride channels}. \emph{Biophys. J.} \textbf{2022},
  \emph{121}, 1336 -- 1347\relax
\mciteBstWouldAddEndPuncttrue
\mciteSetBstMidEndSepPunct{\mcitedefaultmidpunct}
{\mcitedefaultendpunct}{\mcitedefaultseppunct}\relax
\EndOfBibitem
\bibitem[Jiao \latin{et~al.}(2011)Jiao, Leung, Rempe, and
  Nenoff]{jiao2011first}
Jiao,~D.; Leung,~K.; Rempe,~S.~B.; Nenoff,~T.~M. {First Principles Calculations
  of Atomic Nickel Redox Potentials and Dimerization Free Energies: A Study of
  Metal Nanoparticle Growth}. \emph{J. Chem. Theo. Comp.} \textbf{2011},
  \emph{7}, 485--495\relax
\mciteBstWouldAddEndPuncttrue
\mciteSetBstMidEndSepPunct{\mcitedefaultmidpunct}
{\mcitedefaultendpunct}{\mcitedefaultseppunct}\relax
\EndOfBibitem
\bibitem[Marcus(1991)]{ymarc91}
Marcus,~Y. {Thermodynamics of solvation of ions. Part 5.—Gibbs free energy of
  hydration at 298.15 K}. \emph{J. Chem. Soc., Faraday Trans. 1} \textbf{1991},
  \emph{87}, 2995 -- 2999\relax
\mciteBstWouldAddEndPuncttrue
\mciteSetBstMidEndSepPunct{\mcitedefaultmidpunct}
{\mcitedefaultendpunct}{\mcitedefaultseppunct}\relax
\EndOfBibitem
\bibitem[Asthagiri \latin{et~al.}(2003)Asthagiri, Pratt, and
  Ashbaugh]{asthagiri2003jcp}
Asthagiri,~D.; Pratt,~L.~R.; Ashbaugh,~H. Absolute hydration free energies of
  ions, ion--water clusters, and quasichemical theory. \emph{J. Chem. Phys.}
  \textbf{2003}, \emph{119}, 2702--2708\relax
\mciteBstWouldAddEndPuncttrue
\mciteSetBstMidEndSepPunct{\mcitedefaultmidpunct}
{\mcitedefaultendpunct}{\mcitedefaultseppunct}\relax
\EndOfBibitem
\bibitem[Rogers and Rempe(2011)Rogers, and Rempe]{Rogers:2011}
Rogers,~D.~M.; Rempe,~S.~B. {Probing the Thermodynamics of Competitive Ion
  Binding Using Minimum Energy Structures}. \emph{J. Phys. Chem. B}
  \textbf{2011}, \emph{115}, 9116--9129\relax
\mciteBstWouldAddEndPuncttrue
\mciteSetBstMidEndSepPunct{\mcitedefaultmidpunct}
{\mcitedefaultendpunct}{\mcitedefaultseppunct}\relax
\EndOfBibitem
\bibitem[Not()]{Note-1}
The asterisk on \textit{Experiment$^\ast$} of Figure \ref{fig:Exp_QCT_2021}
  emphasizes that these values are not measureable thermodynamically, but are
  inferred from thermodynamic experiments together with extra-thermodynamic
  assumptions. In addition, as noted previously, the Marcus tabulation used
  here\cite{ymarc91} identifies a required standard state adjustment of
  \emph{in}correct sign; See Eq. (4.6) of
  Ref.~\protect\citenum{marcus2015ions}.\relax
\mciteBstWouldAddEndPunctfalse
\mciteSetBstMidEndSepPunct{\mcitedefaultmidpunct}
{}{\mcitedefaultseppunct}\relax
\EndOfBibitem
\bibitem[Varma \latin{et~al.}(2011)Varma, Rogers, Pratt, and
  Rempe]{varma2011design}
Varma,~S.; Rogers,~D.~M.; Pratt,~L.~R.; Rempe,~S.~B. {Design Principles for
  K$^+$ Selectivity in Membrane Transport}. \emph{J. Gen. Phys.} \textbf{2011},
  \emph{137}, 479--488\relax
\mciteBstWouldAddEndPuncttrue
\mciteSetBstMidEndSepPunct{\mcitedefaultmidpunct}
{\mcitedefaultendpunct}{\mcitedefaultseppunct}\relax
\EndOfBibitem
\bibitem[Stevens and Rempe(2016)Stevens, and Rempe]{Stevens:2016}
Stevens,~M.~J.; Rempe,~S. L.~B. {Ion-Specific Effects in Carboxylate Binding
  Sites}. \emph{J. Phys. Chem. B} \textbf{2016}, \emph{120}, 12519--12530\relax
\mciteBstWouldAddEndPuncttrue
\mciteSetBstMidEndSepPunct{\mcitedefaultmidpunct}
{\mcitedefaultendpunct}{\mcitedefaultseppunct}\relax
\EndOfBibitem
\bibitem[Asthagiri \latin{et~al.}(2004)Asthagiri, Pratt, Paulaitis, and
  Rempe]{asthagiri2004hydration}
Asthagiri,~D.; Pratt,~L.~R.; Paulaitis,~M.~E.; Rempe,~S.~B. {Hydration
  Structure and Free Energy of Biomolecularly Specific Aqueous Dications,
  Including Zn$^{2+}$ and First Transition Row Metals}. \emph{J. Am. Chem.
  Soc.} \textbf{2004}, \emph{126}, 1285--1289\relax
\mciteBstWouldAddEndPuncttrue
\mciteSetBstMidEndSepPunct{\mcitedefaultmidpunct}
{\mcitedefaultendpunct}{\mcitedefaultseppunct}\relax
\EndOfBibitem
\bibitem[Chaudhari \latin{et~al.}(2015)Chaudhari, Soniat, and
  Rempe]{Chaudhari:2014wb}
Chaudhari,~M.~I.; Soniat,~M.; Rempe,~S.~B. Octa-Coordination and the Aqueous
  Ba$^{2+}$ Ion. \emph{J. Phys. Chem. B} \textbf{2015}, \emph{119},
  8746--8753\relax
\mciteBstWouldAddEndPuncttrue
\mciteSetBstMidEndSepPunct{\mcitedefaultmidpunct}
{\mcitedefaultendpunct}{\mcitedefaultseppunct}\relax
\EndOfBibitem
\bibitem[Jiao and Rempe(2012)Jiao, and Rempe]{Jiao:2012}
Jiao,~D.; Rempe,~S.~B. {Combined Density Functional Theory (DFT) and Continuum
  Calculations of p$K_a$ in Carbonic Anhydrase}. \emph{Biochem.} \textbf{2012},
  \emph{51}, 5979--5989\relax
\mciteBstWouldAddEndPuncttrue
\mciteSetBstMidEndSepPunct{\mcitedefaultmidpunct}
{\mcitedefaultendpunct}{\mcitedefaultseppunct}\relax
\EndOfBibitem
\bibitem[Dudev and Lim(2013)Dudev, and Lim]{Dudev2013}
Dudev,~T.; Lim,~C. {Importance of Metal Hydration on the Selectivity of
  Mg{$^{2+}$} versus Ca{$^{2+}$} in Magnesium Ion Channels}. \emph{J. Am. Chem.
  Soc.} \textbf{2013}, \emph{135}, 17200--17208\relax
\mciteBstWouldAddEndPuncttrue
\mciteSetBstMidEndSepPunct{\mcitedefaultmidpunct}
{\mcitedefaultendpunct}{\mcitedefaultseppunct}\relax
\EndOfBibitem
\bibitem[Chaudhari and Rempe(2018)Chaudhari, and Rempe]{chaudhari2018SrBa}
Chaudhari,~M.~I.; Rempe,~S.~B. Strontium and Barium in Aqueous Solution and a
  Potassium Channel Binding Site. \emph{J. Chem. Phys.} \textbf{2018},
  \emph{148}, 222831\relax
\mciteBstWouldAddEndPuncttrue
\mciteSetBstMidEndSepPunct{\mcitedefaultmidpunct}
{\mcitedefaultendpunct}{\mcitedefaultseppunct}\relax
\EndOfBibitem
\bibitem[Beck \latin{et~al.}(2006)Beck, Paulaitis, and Pratt]{lrp:book}
Beck,~T.~L.; Paulaitis,~M.~E.; Pratt,~L.~R. \emph{The Potential Distribution
  Theorem and Models of Molecular Solutions}; Cambridge University Press,
  2006\relax
\mciteBstWouldAddEndPuncttrue
\mciteSetBstMidEndSepPunct{\mcitedefaultmidpunct}
{\mcitedefaultendpunct}{\mcitedefaultseppunct}\relax
\EndOfBibitem
\bibitem[Martin \latin{et~al.}(1998)Martin, Hay, and Pratt]{10.1021/jp980229p}
Martin,~R.~L.; Hay,~P.~J.; Pratt,~L.~R. {Hydrolysis of Ferric Ion in Water and
  Conformational Equilibrium}. \emph{J. Phys. Chem. A} \textbf{1998},
  \emph{102}, 3565--3573\relax
\mciteBstWouldAddEndPuncttrue
\mciteSetBstMidEndSepPunct{\mcitedefaultmidpunct}
{\mcitedefaultendpunct}{\mcitedefaultseppunct}\relax
\EndOfBibitem
\bibitem[Rempe \latin{et~al.}(2000)Rempe, Pratt, Hummer, Kress, Martin, and
  Redondo]{Rempe:2000uw}
Rempe,~S.~B.; Pratt,~L.~R.; Hummer,~G.; Kress,~J.~D.; Martin,~R.~L.;
  Redondo,~A. The hydration number of {L}i{$^+$} in liquid water. \emph{J. Am.
  Chem. Soc.} \textbf{2000}, \emph{122}, 966 -- 967\relax
\mciteBstWouldAddEndPuncttrue
\mciteSetBstMidEndSepPunct{\mcitedefaultmidpunct}
{\mcitedefaultendpunct}{\mcitedefaultseppunct}\relax
\EndOfBibitem
\bibitem[Pratt and Asthagiri(2007)Pratt, and Asthagiri]{pratt2007potential}
Pratt,~L.~R.; Asthagiri,~D. \emph{Free Energy Calculations}; Springer, 2007; pp
  323--351\relax
\mciteBstWouldAddEndPuncttrue
\mciteSetBstMidEndSepPunct{\mcitedefaultmidpunct}
{\mcitedefaultendpunct}{\mcitedefaultseppunct}\relax
\EndOfBibitem
\bibitem[Sabo \latin{et~al.}(2008)Sabo, Varma, Martin, and Rempe]{Sabo:h2}
Sabo,~D.; Varma,~S.; Martin,~M.~G.; Rempe,~S.~B. {Studies of the Thermodynamic
  Properties of Hydrogen Gas in Bulk Water}. \emph{J. Phys. Chem. B}
  \textbf{2008}, \emph{112}, 867--876\relax
\mciteBstWouldAddEndPuncttrue
\mciteSetBstMidEndSepPunct{\mcitedefaultmidpunct}
{\mcitedefaultendpunct}{\mcitedefaultseppunct}\relax
\EndOfBibitem
\bibitem[Shah \latin{et~al.}(2007)Shah, Asthagiri, Pratt, and
  Paulaitis]{Shah:2007dm}
Shah,~J.; Asthagiri,~D.; Pratt,~L.; Paulaitis,~M. {Balancing local order and
  long-ranged interactions in the molecular theory of liquid water}. \emph{J.
  Chem. Phys.} \textbf{2007}, \emph{127}, 144508 (1--7)\relax
\mciteBstWouldAddEndPuncttrue
\mciteSetBstMidEndSepPunct{\mcitedefaultmidpunct}
{\mcitedefaultendpunct}{\mcitedefaultseppunct}\relax
\EndOfBibitem
\bibitem[Weber \latin{et~al.}(2011)Weber, Merchant, and
  Asthagiri]{Weber:2011hd}
Weber,~V.; Merchant,~S.; Asthagiri,~D. {Communication: Regularizing binding
  energy distributions and thermodynamics of hydration: Theory and application
  to water modeled with classical and ab initio simulations}. \emph{J. Chem.
  Phys.} \textbf{2011}, \emph{135}, 181101\relax
\mciteBstWouldAddEndPuncttrue
\mciteSetBstMidEndSepPunct{\mcitedefaultmidpunct}
{\mcitedefaultendpunct}{\mcitedefaultseppunct}\relax
\EndOfBibitem
\bibitem[Chempath \latin{et~al.}(2009)Chempath, Pratt, and
  Paulaitis]{chempath2009quasichemical}
Chempath,~S.; Pratt,~L.~R.; Paulaitis,~M.~E. Quasichemical theory with a soft
  cutoff. \emph{J. Chem. Phys.} \textbf{2009}, \emph{130}, 054113\relax
\mciteBstWouldAddEndPuncttrue
\mciteSetBstMidEndSepPunct{\mcitedefaultmidpunct}
{\mcitedefaultendpunct}{\mcitedefaultseppunct}\relax
\EndOfBibitem
\bibitem[Weber and Asthagiri(2012)Weber, and Asthagiri]{Weber:2012kc}
Weber,~V.; Asthagiri,~D. {Regularizing Binding Energy Distributions and the
  Hydration Free Energy of Protein Cytochrome C from All-Atom Simulations}.
  \emph{J. Chem. Theory Comp.} \textbf{2012}, \emph{8}, 3409--3415\relax
\mciteBstWouldAddEndPuncttrue
\mciteSetBstMidEndSepPunct{\mcitedefaultmidpunct}
{\mcitedefaultendpunct}{\mcitedefaultseppunct}\relax
\EndOfBibitem
\bibitem[Tomar \latin{et~al.}(2020)Tomar, Paulaitis, Pratt, and
  Asthagiri]{tomar:jpcl20}
Tomar,~D.~S.; Paulaitis,~M.~E.; Pratt,~L.~R.; Asthagiri,~D.~N. {Hydrophilic
  Interactions Dominate the Inverse Temperature Dependence of Polypeptide
  Hydration Free Energies At\ tributed to Hydrophobicity}. \emph{J. Phys. Chem.
  Lett.} \textbf{2020}, \emph{11}, 9965--9970\relax
\mciteBstWouldAddEndPuncttrue
\mciteSetBstMidEndSepPunct{\mcitedefaultmidpunct}
{\mcitedefaultendpunct}{\mcitedefaultseppunct}\relax
\EndOfBibitem
\bibitem[Chaudhari \latin{et~al.}(2017)Chaudhari, Pratt, and
  Rempe]{Chaudhari:2017gsa}
Chaudhari,~M.~I.; Pratt,~L.~R.; Rempe,~S.~B. {Utility of chemical computations
  in predicting solution free energies of metal ions}. \emph{Mol. Simul.}
  \textbf{2017}, \emph{492}, 1--7\relax
\mciteBstWouldAddEndPuncttrue
\mciteSetBstMidEndSepPunct{\mcitedefaultmidpunct}
{\mcitedefaultendpunct}{\mcitedefaultseppunct}\relax
\EndOfBibitem
\bibitem[Tomasi \latin{et~al.}(2005)Tomasi, Mennucci, and Cammi]{Tomasi:2005tc}
Tomasi,~J.; Mennucci,~B.; Cammi,~R. {Quantum Mechanical Continuum Solvation
  Models}. \emph{Chem. Rev.} \textbf{2005}, \emph{105}, 2999--3093\relax
\mciteBstWouldAddEndPuncttrue
\mciteSetBstMidEndSepPunct{\mcitedefaultmidpunct}
{\mcitedefaultendpunct}{\mcitedefaultseppunct}\relax
\EndOfBibitem
\bibitem[Rempe \latin{et~al.}(2004)Rempe, Asthagiri, and Pratt]{rempe2004inner}
Rempe,~S.~B.; Asthagiri,~D.; Pratt,~L.~R. Inner shell definition and absolute
  hydration free energy of {K$^+$(aq)} on the basis of quasi-chemical theory
  and ab initio molecular dynamics. \emph{Phys. Chem. Chem. Phys.}
  \textbf{2004}, \emph{6}, 1966--1969\relax
\mciteBstWouldAddEndPuncttrue
\mciteSetBstMidEndSepPunct{\mcitedefaultmidpunct}
{\mcitedefaultendpunct}{\mcitedefaultseppunct}\relax
\EndOfBibitem
\bibitem[Sabo \latin{et~al.}(2013)Sabo, Jiao, Varma, Pratt, and
  Rempe]{Sabo:2013gs}
Sabo,~D.; Jiao,~D.; Varma,~S.; Pratt,~L.~R.; Rempe,~S.~B. {Case Study of
  Rb$^+$(aq), Quasi-Chemical Theory of Ion Hydration, and the No Split
  Occupancies Rule}. \emph{Ann. Rep. Prog. Chem, Sect. C (Phys. Chem.)}
  \textbf{2013}, \emph{109}, 266--278\relax
\mciteBstWouldAddEndPuncttrue
\mciteSetBstMidEndSepPunct{\mcitedefaultmidpunct}
{\mcitedefaultendpunct}{\mcitedefaultseppunct}\relax
\EndOfBibitem
\bibitem[Gomez(2021)]{gomez2021rough}
Gomez,~D.~T. Rough Energy-Landscapes of Hydrated Anions: Treatment Using High
  Through-Put Computations. Ph.D.\ thesis, Tulane University School of Science
  and Engineering, 2021\relax
\mciteBstWouldAddEndPuncttrue
\mciteSetBstMidEndSepPunct{\mcitedefaultmidpunct}
{\mcitedefaultendpunct}{\mcitedefaultseppunct}\relax
\EndOfBibitem
\bibitem[Tissandier \latin{et~al.}(1998)Tissandier, Cowen, Feng, Gundlach,
  Cohen, Earhart, Coe, and Tuttle]{tissandier1998proton}
Tissandier,~M.~D.; Cowen,~K.~A.; Feng,~W.~Y.; Gundlach,~E.; Cohen,~M.~H.;
  Earhart,~A.~D.; Coe,~J.~V.; Tuttle,~T.~R. The Proton's Absolute Aqueous
  Enthalpy and Gibbs Free Energy of Solvation From Cluster-Ion Solvation Data.
  \emph{J. Phys. Chem. A} \textbf{1998}, \emph{102}, 7787--7794\relax
\mciteBstWouldAddEndPuncttrue
\mciteSetBstMidEndSepPunct{\mcitedefaultmidpunct}
{\mcitedefaultendpunct}{\mcitedefaultseppunct}\relax
\EndOfBibitem
\bibitem[Basdogan \latin{et~al.}(2020)Basdogan, Groenenboom, Henderson, De,
  Rempe, and Keith]{basdogan}
Basdogan,~Y.; Groenenboom,~M.~C.; Henderson,~E.; De,~S.; Rempe,~S.~B.;
  Keith,~J.~A. Machine learning-guided approach for studying solvation
  environments. \emph{J. Chem. Theory Comput.} \textbf{2020}, \emph{16}, 633 --
  42\relax
\mciteBstWouldAddEndPuncttrue
\mciteSetBstMidEndSepPunct{\mcitedefaultmidpunct}
{\mcitedefaultendpunct}{\mcitedefaultseppunct}\relax
\EndOfBibitem
\bibitem[Heuft and Meijer(2003)Heuft, and Meijer]{Heuft:2003iva}
Heuft,~J.~M.; Meijer,~E.~J. {Density functional theory based molecular-dynamics
  study of aqueous chloride solvation}. \emph{J. Chem. Phys.} \textbf{2003},
  \emph{119}, 11788 -- 11791\relax
\mciteBstWouldAddEndPuncttrue
\mciteSetBstMidEndSepPunct{\mcitedefaultmidpunct}
{\mcitedefaultendpunct}{\mcitedefaultseppunct}\relax
\EndOfBibitem
\bibitem[Heuft and Meijer(2005)Heuft, and Meijer]{Heuft:2005jt}
Heuft,~J.~M.; Meijer,~E.~J. {Density functional theory based molecular-dynamics
  study of aqueous iodide solvation}. \emph{J. Chem. Phys.} \textbf{2005},
  \emph{123}, 094506 -- 6\relax
\mciteBstWouldAddEndPuncttrue
\mciteSetBstMidEndSepPunct{\mcitedefaultmidpunct}
{\mcitedefaultendpunct}{\mcitedefaultseppunct}\relax
\EndOfBibitem
\bibitem[Heuft and Meijer(2005)Heuft, and Meijer]{Heuft:2005kx}
Heuft,~J.~M.; Meijer,~E.~J. {Density functional theory based molecular-dynamics
  study of aqueous fluoride solvation}. \emph{J. Chem. Phys.} \textbf{2005},
  \emph{122}, 094501 -- 8\relax
\mciteBstWouldAddEndPuncttrue
\mciteSetBstMidEndSepPunct{\mcitedefaultmidpunct}
{\mcitedefaultendpunct}{\mcitedefaultseppunct}\relax
\EndOfBibitem
\bibitem[Wiktor \latin{et~al.}(2017)Wiktor, Bruneval, and
  Pasquarello]{wiktor2017partial}
Wiktor,~J.; Bruneval,~F.; Pasquarello,~A. Partial Molar Volumes of Aqua Ions
  from First Principles. \emph{J. Chem. Theory Comput.} \textbf{2017},
  \emph{13}, 3427--3431\relax
\mciteBstWouldAddEndPuncttrue
\mciteSetBstMidEndSepPunct{\mcitedefaultmidpunct}
{\mcitedefaultendpunct}{\mcitedefaultseppunct}\relax
\EndOfBibitem
\bibitem[Duignan \latin{et~al.}(2017)Duignan, Baer, Schenter, and
  Mundy]{Duignan:2017iha}
Duignan,~T.~T.; Baer,~M.~D.; Schenter,~G.~K.; Mundy,~C.~J. {Real single ion
  solvation free energies with quantum mechanical simulation}. \emph{Chem.
  Sci.} \textbf{2017}, \emph{8}, 6131--6140\relax
\mciteBstWouldAddEndPuncttrue
\mciteSetBstMidEndSepPunct{\mcitedefaultmidpunct}
{\mcitedefaultendpunct}{\mcitedefaultseppunct}\relax
\EndOfBibitem
\bibitem[Duignan \latin{et~al.}(2021)Duignan, Kathmann, Schenter, and
  Mundy]{Duignan2021}
Duignan,~T.~T.; Kathmann,~S.~M.; Schenter,~G.~K.; Mundy,~C.~J. {Toward a
  First-Principles Framework for Predicting Collective Properties of
  Electrolytes}. \emph{Acc. Chem. Res.} \textbf{2021}, \emph{54},
  2833--2843\relax
\mciteBstWouldAddEndPuncttrue
\mciteSetBstMidEndSepPunct{\mcitedefaultmidpunct}
{\mcitedefaultendpunct}{\mcitedefaultseppunct}\relax
\EndOfBibitem
\bibitem[Kresse and Furthm{\"u}ller(1996)Kresse, and
  Furthm{\"u}ller]{kresse1996efficient}
Kresse,~G.; Furthm{\"u}ller,~J. Efficient Iterative Schemes for Ab Initio
  Total-Energy Calculations Using a Plane-Wave Basis Set. \emph{Phys. Rev. B}
  \textbf{1996}, \emph{54}, 11169\relax
\mciteBstWouldAddEndPuncttrue
\mciteSetBstMidEndSepPunct{\mcitedefaultmidpunct}
{\mcitedefaultendpunct}{\mcitedefaultseppunct}\relax
\EndOfBibitem
\bibitem[Kühne \latin{et~al.}(2020)Kühne, Iannuzzi, Ben, Rybkin, Seewald,
  Stein, Laino, Khaliullin, Schütt, Schiffmann, Golze, Wilhelm, Chulkov,
  Bani-Hashemian, Weber, Borštnik, Taillefumier, Jakobovits, Lazzaro, Pabst,
  Müller, Schade, Guidon, Andermatt, Holmberg, Schenter, Hehn, Bussy,
  Belleflamme, Tabacchi, Glöß, Lass, Bethune, Mundy, Plessl, Watkins,
  VandeVondele, Krack, and Hutter]{10.1063/5.0007045}
Kühne,~T.~D.; Iannuzzi,~M.; Ben,~M.~D.; Rybkin,~V.~V.; Seewald,~P.; Stein,~F.;
  Laino,~T.; Khaliullin,~R.~Z.; Schütt,~O.; Schiffmann,~F.; Golze,~D.;
  Wilhelm,~J.; Chulkov,~S.; Bani-Hashemian,~M.~H.; Weber,~V.; Borštnik,~U.;
  Taillefumier,~M.; Jakobovits,~A.~S.; Lazzaro,~A.; Pabst,~H.; Müller,~T.;
  Schade,~R.; Guidon,~M.; Andermatt,~S.; Holmberg,~N.; Schenter,~G.~K.;
  Hehn,~A.; Bussy,~A.; Belleflamme,~F.; Tabacchi,~G.; Glöß,~A.; Lass,~M.;
  Bethune,~I.; Mundy,~C.~J.; Plessl,~C.; Watkins,~M.; VandeVondele,~J.;
  Krack,~M.; Hutter,~J. {CP2K: An electronic structure and molecular dynamics
  software package - Quickstep: Efficient and accurate electronic structure
  calculations}. \emph{J. Chem. Phys.} \textbf{2020}, \emph{152}, 194103\relax
\mciteBstWouldAddEndPuncttrue
\mciteSetBstMidEndSepPunct{\mcitedefaultmidpunct}
{\mcitedefaultendpunct}{\mcitedefaultseppunct}\relax
\EndOfBibitem
\bibitem[Goedecker \latin{et~al.}(1996)Goedecker, Teter, and
  Hutter]{goedecker1996separable}
Goedecker,~S.; Teter,~M.; Hutter,~J. Separable Dual-Space {G}aussian
  Pseudopotentials. \emph{Phys. Rev. B} \textbf{1996}, \emph{54}, 1703\relax
\mciteBstWouldAddEndPuncttrue
\mciteSetBstMidEndSepPunct{\mcitedefaultmidpunct}
{\mcitedefaultendpunct}{\mcitedefaultseppunct}\relax
\EndOfBibitem
\bibitem[Lippert \latin{et~al.}(1999)Lippert, Hutter, and
  Parrinello]{lippert1999gaussian}
Lippert,~G.; Hutter,~J.; Parrinello,~M. The {G}aussian and Augmented-Plane-Wave
  Density Functional Method for Ab Initio Molecular Dynamics Simulations.
  \emph{Theo. Chem. Accts.} \textbf{1999}, \emph{103}, 124--140\relax
\mciteBstWouldAddEndPuncttrue
\mciteSetBstMidEndSepPunct{\mcitedefaultmidpunct}
{\mcitedefaultendpunct}{\mcitedefaultseppunct}\relax
\EndOfBibitem
\bibitem[Nos{\'e}(1984)]{NOSE}
Nos{\'e},~S. {A Molecular Dynamics Method for Simulations in the Canonical
  Ensemble}. \emph{Mol. Phys.} \textbf{1984}, \emph{52}, 255--268\relax
\mciteBstWouldAddEndPuncttrue
\mciteSetBstMidEndSepPunct{\mcitedefaultmidpunct}
{\mcitedefaultendpunct}{\mcitedefaultseppunct}\relax
\EndOfBibitem
\bibitem[Frisch \latin{et~al.}(2016)Frisch, Trucks, Schlegel, Scuseria, Robb,
  Cheeseman, Scalmani, Barone, Petersson, Nakatsuji, Li, Caricato, Marenich,
  Bloino, Janesko, Gomperts, Mennucci, Hratchian, Ortiz, Izmaylov, Sonnenberg,
  Williams-Young, Ding, Lipparini, Egidi, Goings, Peng, Petrone, Henderson,
  Ranasinghe, Zakrzewski, Gao, Rega, Zheng, Liang, Hada, Ehara, Toyota, Fukuda,
  Hasegawa, Ishida, Nakajima, Honda, Kitao, Nakai, Vreven, Throssell,
  Montgomery, Peralta, Ogliaro, Bearpark, Heyd, Brothers, Kudin, Staroverov,
  Keith, Kobayashi, Normand, Raghavachari, Rendell, Burant, Iyengar, Tomasi,
  Cossi, Millam, Klene, Adamo, Cammi, Ochterski, Martin, Morokuma, Farkas,
  Foresman, and Fox]{g16}
Frisch,~M.~J.; Trucks,~G.~W.; Schlegel,~H.~B.; Scuseria,~G.~E.; Robb,~M.~A.;
  Cheeseman,~J.~R.; Scalmani,~G.; Barone,~V.; Petersson,~G.~A.; Nakatsuji,~H.;
  Li,~X.; Caricato,~M.; Marenich,~A.~V.; Bloino,~J.; Janesko,~B.~G.;
  Gomperts,~R.; Mennucci,~B.; Hratchian,~H.~P.; Ortiz,~J.~V.; Izmaylov,~A.~F.;
  Sonnenberg,~J.~L.; Williams-Young,~D.; Ding,~F.; Lipparini,~F.; Egidi,~F.;
  Goings,~J.; Peng,~B.; Petrone,~A.; Henderson,~T.; Ranasinghe,~D.;
  Zakrzewski,~V.~G.; Gao,~J.; Rega,~N.; Zheng,~G.; Liang,~W.; Hada,~M.;
  Ehara,~M.; Toyota,~K.; Fukuda,~R.; Hasegawa,~J.; Ishida,~M.; Nakajima,~T.;
  Honda,~Y.; Kitao,~O.; Nakai,~H.; Vreven,~T.; Throssell,~K.;
  Montgomery,~J.~A.,~{Jr.}; Peralta,~J.~E.; Ogliaro,~F.; Bearpark,~M.~J.;
  Heyd,~J.~J.; Brothers,~E.~N.; Kudin,~K.~N.; Staroverov,~V.~N.; Keith,~T.~A.;
  Kobayashi,~R.; Normand,~J.; Raghavachari,~K.; Rendell,~A.~P.; Burant,~J.~C.;
  Iyengar,~S.~S.; Tomasi,~J.; Cossi,~M.; Millam,~J.~M.; Klene,~M.; Adamo,~C.;
  Cammi,~R.; Ochterski,~J.~W.; Martin,~R.~L.; Morokuma,~K.; Farkas,~O.;
  Foresman,~J.~B.; Fox,~D.~J. Gaussian~16 {R}evision {C}.01. 2016; Gaussian
  Inc. Wallingford CT\relax
\mciteBstWouldAddEndPuncttrue
\mciteSetBstMidEndSepPunct{\mcitedefaultmidpunct}
{\mcitedefaultendpunct}{\mcitedefaultseppunct}\relax
\EndOfBibitem
\bibitem[Pratt \latin{et~al.}(1997)Pratt, Tawa, Hummer, Garc{\'\i}a, and
  Corcelli]{pratt1997boundary}
Pratt,~L.~R.; Tawa,~G.~J.; Hummer,~G.; Garc{\'\i}a,~A.~E.; Corcelli,~S.~A.
  Boundary Integral Methods for the Poisson Equation of Continuum Dielectric
  Solvation Models. \emph{Int. J. Quant. Chem.} \textbf{1997}, \emph{64},
  121--141\relax
\mciteBstWouldAddEndPuncttrue
\mciteSetBstMidEndSepPunct{\mcitedefaultmidpunct}
{\mcitedefaultendpunct}{\mcitedefaultseppunct}\relax
\EndOfBibitem
\bibitem[Linder and Hoernschemeyer(1967)Linder, and
  Hoernschemeyer]{linder1967cavity}
Linder,~B.; Hoernschemeyer,~D. {Cavity Concept in Dielectric Theory}. \emph{J.
  Chem. Phys.} \textbf{1967}, \emph{46}, 784\relax
\mciteBstWouldAddEndPuncttrue
\mciteSetBstMidEndSepPunct{\mcitedefaultmidpunct}
{\mcitedefaultendpunct}{\mcitedefaultseppunct}\relax
\EndOfBibitem
\bibitem[Asthagiri \latin{et~al.}(2003)Asthagiri, Pratt, Kress, and
  Gomez]{AsthagiriD:ThehsH}
Asthagiri,~D.; Pratt,~L.~R.; Kress,~J.; Gomez,~M.~A. {The Hydration State of
  HO$^-$(aq).} \emph{Chem. Phys. Lett.} \textbf{2003}, \emph{380},
  530--535\relax
\mciteBstWouldAddEndPuncttrue
\mciteSetBstMidEndSepPunct{\mcitedefaultmidpunct}
{\mcitedefaultendpunct}{\mcitedefaultseppunct}\relax
\EndOfBibitem
\bibitem[Merchant \latin{et~al.}(2011)Merchant, Dixit, Dean, and
  Asthagiri]{Merchant:2011ga}
Merchant,~S.; Dixit,~P.~D.; Dean,~K.~R.; Asthagiri,~D. {Ion-water Clusters,
  Bulk Medium Effects, and Ion Hydration}. \emph{J. Chem. Phys.} \textbf{2011},
  \emph{135}, 54505\relax
\mciteBstWouldAddEndPuncttrue
\mciteSetBstMidEndSepPunct{\mcitedefaultmidpunct}
{\mcitedefaultendpunct}{\mcitedefaultseppunct}\relax
\EndOfBibitem
\bibitem[Marcus(1994)]{Marcus:1994ci}
Marcus,~Y. {A Simple Empirical Model Describing the Thermodynamics of Hydration
  of Ions of Widely Varying Charges, Sizes, and Shapes}. \emph{Biophys. Chem.}
  \textbf{1994}, \emph{51}, 111--127\relax
\mciteBstWouldAddEndPuncttrue
\mciteSetBstMidEndSepPunct{\mcitedefaultmidpunct}
{\mcitedefaultendpunct}{\mcitedefaultseppunct}\relax
\EndOfBibitem
\bibitem[Ohtaki and Radnai(1993)Ohtaki, and Radnai]{Ohtaki1993}
Ohtaki,~H.; Radnai,~T. {Structure and Dynamics of Hydrated Ions}. \emph{Chem.
  Rev.} \textbf{1993}, \emph{93}, 1157--1204\relax
\mciteBstWouldAddEndPuncttrue
\mciteSetBstMidEndSepPunct{\mcitedefaultmidpunct}
{\mcitedefaultendpunct}{\mcitedefaultseppunct}\relax
\EndOfBibitem
\bibitem[Megyes \latin{et~al.}(2008)Megyes, Balint, Grosz, Radnai, Bako, and
  Sipos]{Sipos:2008}
Megyes,~T.; Balint,~S.; Grosz,~T.; Radnai,~T.; Bako,~I.; Sipos,~P. {The
  Structure of Aqueous Sodium Hydroxide Solutions: A Combined Solution X-ray
  Diffraction and Simulation Study}. \emph{J. Chem. Phys} \textbf{2008},
  \emph{128}, 044501\relax
\mciteBstWouldAddEndPuncttrue
\mciteSetBstMidEndSepPunct{\mcitedefaultmidpunct}
{\mcitedefaultendpunct}{\mcitedefaultseppunct}\relax
\EndOfBibitem
\bibitem[Fulton \latin{et~al.}(2010)Fulton, Schenter, Baer, Mundy, Dang, and
  Balasubramanian]{Fulton:I}
Fulton,~J.~L.; Schenter,~G.~K.; Baer,~M.~D.; Mundy,~C.~J.; Dang,~L.~X.;
  Balasubramanian,~M. {Probing the Hydration Structure of Polarizable Halides:
  A Multiedge XAFS and Molecular Dynamics Study of the Iodide Anion }. \emph{J.
  Phys. Chem. B} \textbf{2010}, \emph{114}, 12926 -- 12937\relax
\mciteBstWouldAddEndPuncttrue
\mciteSetBstMidEndSepPunct{\mcitedefaultmidpunct}
{\mcitedefaultendpunct}{\mcitedefaultseppunct}\relax
\EndOfBibitem
\bibitem[Rogers and Beck(2010)Rogers, and Beck]{Rogers:2010gh}
Rogers,~D.~M.; Beck,~T.~L. {Quasichemical and Structural Analysis of
  Polarizable Anion Hydration}. \emph{J. Chem. Phys.} \textbf{2010},
  \emph{132}, 014505--13\relax
\mciteBstWouldAddEndPuncttrue
\mciteSetBstMidEndSepPunct{\mcitedefaultmidpunct}
{\mcitedefaultendpunct}{\mcitedefaultseppunct}\relax
\EndOfBibitem
\bibitem[Marcus(2015)]{marcus2015ions}
Marcus,~Y. \emph{Ions in Solution and their Solvation}; John Wiley \& Sons,
  2015; Note Eq. (4.6).\relax
\mciteBstWouldAddEndPunctfalse
\mciteSetBstMidEndSepPunct{\mcitedefaultmidpunct}
{}{\mcitedefaultseppunct}\relax
\EndOfBibitem
\end{mcitethebibliography}

\providecommand{\latin}[1]{#1}
\makeatletter
\providecommand{\doi}
  {\begingroup\let\do\@makeother\dospecials
  \catcode`\{=1 \catcode`\}=2 \doi@aux}
\providecommand{\doi@aux}[1]{\endgroup\texttt{#1}}
\makeatother
\providecommand*\mcitethebibliography{\thebibliography}
\csname @ifundefined\endcsname{endmcitethebibliography}
  {\let\endmcitethebibliography\endthebibliography}{}

\newpage\pagebreak
\begin{figure*}[ht]
\includegraphics[width=\textwidth]{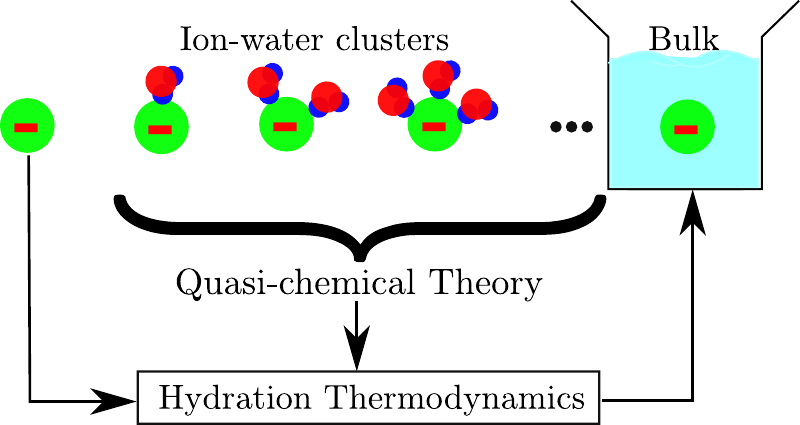}
\end{figure*}

\end{document}